\documentclass[a4paper,11pt]{article}
\pdfoutput=1 
\usepackage{jcappub, bm, color} 
\usepackage{amssymb,amsfonts,slashed,amsthm,amsmath,graphicx, soul, empheq, cancel}
\usepackage[caption=false]{subfig}
\begin{document}

\newcommand{\vev}[1]{ \left\langle {#1} \right\rangle }
\newcommand{\bra}[1]{ \langle {#1} | }
\newcommand{\ket}[1]{ | {#1} \rangle }
\newcommand{\eV}{ \ {\rm eV} }
\newcommand{\KeV}{ \ {\rm keV} }
\newcommand{\MeV}{\  {\rm MeV} }
\newcommand{\GeV}{\  {\rm GeV} }
\newcommand{\TeV}{\  {\rm TeV} }
\newcommand{\1}{\mbox{1}\hspace{-0.25em}\mbox{l}}
\newcommand{\Red}[1]{{\color{red} {#1}}}

\newcommand{\lmk}{\left(}  
\newcommand{\rmk}{\right)}
\newcommand{\lkk}{\left[}  
\newcommand{\rkk}{\right]}
\newcommand{\lhk}{\left \{ }  
\newcommand{\rhk}{\right \} }
\newcommand{\del}{\partial}  
\newcommand{\la}{\left\langle} 
\newcommand{\ra}{\right\rangle}
\newcommand{\half}{\frac{1}{2}}

\newcommand{\bea}{\begin{array}}
\newcommand{\eea}{\end{array}}
\newcommand{\beq}{\begin{eqnarray}}
\newcommand{\eeq}{\end{eqnarray}}
\newcommand{\eq}[1]{Eq.~(\ref{#1})}

\newcommand{\dd}{\mathrm{d}}
\newcommand{\Mpl}{M_{\rm Pl}}
\newcommand{\mg}{m_{3/2}}
\newcommand{\abs}[1]{\left\vert {#1} \right\vert}
\newcommand{\mphi}{m_{\phi}}
\newcommand{\Hz}{\ {\rm Hz}}
\newcommand{\for}{\quad \text{for }}
\newcommand{\Min}{\text{Min}}
\newcommand{\Max}{\text{Max}}
\newcommand{\Kahler}{K\"{a}hler }
\newcommand{\cphi}{\varphi}
\newcommand{\Tr}{\text{Tr}}
\newcommand{\diag}{{\rm diag}}

\newcommand{\SUf}{SU(3)_{\rm f}}
\newcommand{\Upq}{U(1)_{\rm PQ}}
\newcommand{\Zpq}{Z^{\rm PQ}_3}
\newcommand{\Cpq}{C_{\rm PQ}}
\newcommand{\ubar}{u^c}
\newcommand{\dbar}{d^c}
\newcommand{\ebar}{e^c}
\newcommand{\nubar}{\nu^c}
\newcommand{\Ndw}{N_{\rm DW}}
\newcommand{\Fpq}{F_{\rm PQ}}
\newcommand{\fpq}{v_{\rm PQ}}
\newcommand{\Br}{{\rm Br}}
\newcommand{\Lag}{\mathcal{L}}
\newcommand{\Lqcd}{\Lambda_{\rm QCD}}

\newcommand{\ji}{j_{\rm inf}} 
\newcommand{\jb}{j_{B-L}} 
\newcommand{\M}{M} 
\newcommand{\im}{{\rm Im} }
\newcommand{\re}{{\rm Re} }

\def\lrf#1#2{ \left(\frac{#1}{#2}\right)}
\def\lrfp#1#2#3{ \left(\frac{#1}{#2} \right)^{#3}}
\def\lrp#1#2{\left( #1 \right)^{#2}}
\def\REF#1{Ref.~\cite{#1}}
\def\SEC#1{Sec.~\ref{#1}}
\def\FIG#1{Fig.~\ref{#1}}
\def\EQ#1{Eq.~(\ref{#1})}
\def\EQS#1{Eqs.~(\ref{#1})}
\def\TEV#1{10^{#1}{\rm\,TeV}}
\def\GEV#1{10^{#1}{\rm\,GeV}}
\def\MEV#1{10^{#1}{\rm\,MeV}}
\def\KEV#1{10^{#1}{\rm\,keV}}
\def\blue#1{\textcolor{blue}{#1}}
\def\red#1{\textcolor{blue}{#1}}

\newcommand{\eff}{\Delta N_{\rm eff}}
\newcommand{\neff}{\Delta N_{\rm eff}}
\newcommand{\cc}{\Omega_\Lambda}
\newcommand{\Mpc}{\ {\rm Mpc}}
\newcommand{\Msolar}{M_\odot}

\def\sn#1{\textcolor{red}{#1}}
\def\SN#1{\textcolor{red}{[{\bf SN:} #1]}}
\def\ft#1{\textcolor{magenta}{#1}}
\def\FT#1{\textcolor{magenta}{[{\bf FT:} #1]}}
\def\my#1{\textcolor{blue}{#1}}
\def\MY#1{\textcolor{blue}{[{\bf MY:} #1]}}

%%%%%%%%%%%%%%%%%%%%%%%%%%%%%%%%%%%%%%%%%%%%%%%%%%%%%%%%%%%%%%%
%######################
\begin{flushright}
TU-1211
\end{flushright}
%######################

\title{
Dissipation of axion energy 
\\
via the Schwinger and Witten effects
}

\author{
Kwang Sik Jeong,$^1$
}
\author{
Shota Nakagawa,$^{2,3,4}$
}
\author{
Fuminobu Takahashi,$^2$
}
\author{
Masaki Yamada$^{2,5}$
}

\affiliation{$^1$ Department of Physics, Pusan National University, Busan 46241, Korea} 

\affiliation{$^2$ Department of Physics, Tohoku University, Sendai, Miyagi 980-8578, Japan} 

\affiliation{$^3$ Tsung-Dao Lee Institute, Shanghai Jiao Tong University, \\
520 Shengrong Road, Shanghai 201210, China} 

\affiliation{$^4$ School of Physics and Astronomy, Shanghai Jiao Tong University, \\
800 Dongchuan Road, Shanghai 200240, China} 

\affiliation{$^5$ FRIS, Tohoku University, Sendai, Miyagi 980-8578, Japan}

\abstract{
In the presence of an anomalous CP phase in a U(1) gauge theory, a monopole becomes a dyon via the Witten effect. When the anomalous CP phase is promoted to a dynamical field, the axion, the electric charge of the dyon changes according to the coherent motion of the axion oscillation. Once the electric charge exceeds a certain threshold, the Schwinger pair production of charged particles becomes efficient near the surface of the dyon. 
These non-perturbative effects lead to the back reaction of the axion dynamics by causing the dissipation of the axion oscillation energy and the change of the effective potential due to the Witten effect.
Taking these effects into account, we consider the dynamics of the whole system, including the axion, monopole, and charged heavy vector bosons, and discuss to what extent the axion abundance is modified.
We also discuss the electric dipole radiation from a bound state of a monopole-anti-monopole pair due to the axion coherent oscillations. 
}

\emailAdd{ksjeong@pusan.ac.kr}
\emailAdd{shota.nakagawa.d7@tohoku.ac.jp}
\emailAdd{fumi@tohoku.ac.jp}
\emailAdd{m.yamada@tohoku.ac.jp}

\maketitle
\flushbottom

%%%%%%%%%%%%%%%%%%%%%%%%%%%%%%%%%%%%%%%%%%%%%%%%%%%%%%%%%%%
%%%%%%%%%%%%%%%%%%%%%%%%%%%%%%%%%%%%%%%%%%%%%%%%%%%%%%%%%%%
\section{Introduction
\label{introduction}}

The axion, which is a pseudo-Nambu-Goldstone boson associated with the spontaneous breaking of an anomalous global U(1) symmetry, is recognized as a good candidate for dark matter (DM) because of the weakness of its interactions and the longevity. 
In particular, the QCD axion, predicted by the Peccei-Quinn (PQ) mechanism \cite{Peccei:1977hh, Peccei:1977ur, Weinberg:1977ma,Wilczek:1977pj}, represents one of the most promising solutions to the strong CP problem in the Standard Model (SM). 
One can also consider axion-like particles (which we also call axions in this paper), whose mass and decay constant are free parameters. 
They are currently the subject of various observational and experimental studies. See Refs.~\cite{Jaeckel:2010ni,Ringwald:2012hr,Arias:2012az,Graham:2015ouw,Marsh:2015xka,Irastorza:2018dyq, DiLuzio:2020wdo} for reviews.

The axion can be produced by the so-called misalignment mechanism \cite{Preskill:1982cy,Abbott:1982af,Dine:1982ah}, in which it begins to oscillate coherently around the potential minimum when the Hubble parameter becomes comparable to its mass. The resulting axion abundance is determined by its mass, decay constant, and initial misalignment angle. 
A viable parameter space is determined in terms of these parameters to explain the DM abundance.
Since various axion search experiments and astrophysical observations place constraints on these parameters, 
it is important to consider different possibilities for the axion dynamics that may change the parameter space of interest. 
A stochastic axion \cite{Graham:2018jyp,Guth:2018hsa,Ho:2019ayl,Nakagawa:2020eeg,Reig:2021ipa,Murai:2023xjn} and pi-shift inflation \cite{Co:2018mho,Takahashi:2019pqf,Huang:2020etx} are proposed as a scenario to achieve a small or large initial angle with $\theta_{\rm ini} \ll 1$ or $\theta_{\rm ini} \sim \pi$, respectively.

Another scenario for modifying the parameter space involves the dissipation of axion energy into the gauge sector coupled to the axion. For example, if the axion has a relatively strong coupling to an Abelian gauge sector, tachyonic preheating can occur, effectively producing dark photons and reducing the axion abundance~\cite{Agrawal:2017eqm,Kitajima:2017peg}. Even without such a strong coupling, similar dissipation can occur if the onset of axion oscillation is delayed~\cite{Kitajima:2023pby} in the trapped misalignment mechanism~\cite{Higaki:2016yqk,Jeong:2022kdr} (see also Ref.~\cite{Nakagawa:2022wwm}). Similarly, a scenario has recently been proposed where axion abundances are suppressed by a comparatively strong coupling to a hidden non-Abelian gauge sector, which causes the axion energy to dissipate due to the frictional force arising from the sphaleron effect~\cite{Papageorgiou:2022prc,Choi:2022nlt}. In this paper we explore another dissipation process via a coupling between the axion and an Abelian gauge sector.

In general, an axion can gain an effective mass through its anomalous coupling to a gauge group. If the gauge group is non-Abelian, the axion will have an effective potential due to non-perturbative effects. 
On the other hand, if the gauge group is Abelian, 
the vacuum structure of the gauge field is trivial and its anomalous coupling does not generate a mass for the axion. 
However, this is not the case when monopoles are present in the background, as they introduce non-trivial boundary conditions. 
In fact, it is known that the so-called $\Theta$ term becomes physical in the presence of a monopole, which becomes a dyon\footnote{A dyon represents a monopole with an electric charge. Throughout this paper, we use them interchangeably.} and obtains an electric charge proportional to $\Theta$ via the Witten effect~\cite{Witten:1979ey}. In the case of an axion coupled to the Abelian gauge field, 
the parameter $\Theta$ is replaced by the axion field. 
This implies that the electric charge of the dyon is proportional to the axion field value. Since the induced charge generates an electric field around the dyon and  increases the total energy of the system, the axion should feel a backreaction that can be represented as an effective potential~\cite{Fischler:1983sc}. 
Since the density of monopoles is diluted by cosmic expansion, the effective mass for the axion decreases with time. Consequently, the whole system composed of the axion, U(1) gauge field, and monopoles, evolves in a non-trivial way. The impact of the Witten effect on axion dynamics has been studied in terms of suppression of the axion abundance and isocurvature fluctuations~\cite{Kawasaki:2015lpf,Nomura:2015xil,Kawasaki:2017xwt,Nakagawa:2020zjr}. In this paper, we discuss a new dissipative process induced by monopoles that has not been considered before.

We explore the possibility that an axion  dissipates its energy through the anomalous coupling to an Abelian gauge sector. 
Specifically, we consider a model 
with a 't Hooft-Polyakov monopole and heavy vector gauge bosons $W^\pm$ in the gauge sector, assuming that the U(1) gauge symmetry arises from a spontaneous symmetry breaking of the SU(2) gauge symmetry by an adjoint Higgs field. 
An axion is assumed to couple to the U(1) gauge field. 
The monopole then becomes a dyon with an electric charge proportional to the value of the axion field. 
When this electric charge exceeds a certain threshold, the electric field near the surface of the dyon becomes strong enough to produce pairs of $W^\pm$ bosons via the Schwinger effect. The gauge boson with the opposite electric charge to the dyon is expected to be absorbed by the dyon and reduce its electric charge. The one with the same charge is expected to move away. This process can be interpreted as the decay of a dyon with charge $Q (>0)$ into a dyon with charge $Q-1$ and a charged gauge boson $W^+$. As a result, some part of the axion's energy is dissipated through the production of charged gauge bosons.
The dissipation process becomes effective when the electric charge of the dyon exceeds the threshold for $W^\pm$ pair production on the surface of the dyon. This is indeed achieved when there is a substantial hierarchy between the decay constant, which defines the coupling between the U(1) gauge field and the axion, and the amplitude of the axion oscillations. 
One may be able to interpret this dissipation effect as the Abelian counterpart or reminiscent of the energy dissipation by the sphaleron effect~\cite{Papageorgiou:2022prc,Choi:2022nlt} 
because the interior of the 't Hooft-Polyakov monopole preserves the information of the non-Abelian gauge sector. 
In the following sections, we will explore this new dissipation effect and discuss the dynamics of the axion, the monopole, and the charged particles. We will show that this new dissipative effect of the monopole could significantly change the abundance of the axion.

Specifically, one must decide which case to compare in order to say whether the axion abundance is reduced or enhanced.
Let us consider the case without the Schwinger effect.
If the dyons do not decay by emitting the $W$ boson, the axion  oscillates around the minimum of the potential induced
by the Witten effect. In this scenario, the additional particle production
is known to be exponentially suppressed by the adiabatic suppression mechanism~\cite{Kawasaki:2015lpf,Kawasaki:2017xwt,Nakagawa:2020zjr}.
However, in the case of our primary interest, the axion potential changes due to the decay of dyons, leading to a slight deviation of axion field from the temporal minimum of the potential.
This correspond to the extra particle production resulting from the violation of the adiabatic condition. Thus, compared to the case without  the Schwinger effect, the axion abundance is
enhanced. On the other hand, compared to the case without the Witten effect, we observe a suppression of the axion abundance due to the dissipation effect via the Schwinger effect. In this paper, we will mainly consider the axion abundance from the latter viewpoint.

The remainder of this paper is organized as follows. In Sec.~\ref{sec:dissipation}, we explain the microscopic aspect of the dissipation process in our model. In Sec.~\ref{sec:dynamics0}, 
we consider the backreaction of the Schwinger process under the axion dynamics. 
In Sec.~\ref{sec:evolution}, we numerically study the cosmological evolution and calculate the axion abundance. The last section is devoted to discussion and conclusions. In Appendix, we discuss another dissipation effect due to the electric dipole radiation from the monopole-anti-monopole pairs, which would be relevant in some parameter space.

%%%%%%%%%%%%%%%%%%%%%%%%%%%%%%%%%%%%%%%%%%%%%%%%%%%%%%%%%%%
%%%%%%%%%%%%%%%%%%%%%%%%%%%%%%%%%%%%%%%%%%%%%%%%%%%%%%%%%%%
\section{Microscopic aspects of the dissipation processes
\label{sec:dissipation}}
The Chern-Simons term in an Abelian gauge theory induces an electric charge on a magnetic monopole. This is known as the Witten effect \cite{Witten:1979ey}. This implies that the axion coupling to the gauge field induces a dynamical electric charge of the monopole \cite{Fischler:1983sc}. However, a monopole with a sufficiently large electric charge becomes unstable and produces light charged particles via the Schwinger effect around its surface~\cite{Heisenberg:1936nmg,Schwinger:1951nm}. 
The produced charged particles may or may not be absorbed into the monopole or annihilate subsequently. 
We study the energy dissipation of the axion due to the Schwinger effect in conjunction with the Witten effect in the presence of the hidden magnetic monopole. 
In this section we first specify the setup and scenario we consider in Sec.~\ref{sec:setup}, 
review the Witten effect in Sec.~\ref{sec:witten}, 
explain the phenomenology of the Schwinger effect in Sec.~\ref{sec:dyon}, 
and 
consider the absorption and annihilation of the produced charged particles in Sec.~\ref{sec:Wboson}.

\subsection{Setup}
\label{sec:setup}
First we describe the setup of our scenario. We are interested in an Abelian gauge theory containing a monopole, a charged heavy particle, and an axion. Although it is not necessary to assume a specific UV theory for our purpose, for concreteness we consider that the monopole and the heavy particle arise from a spontaneous symmetry breaking of a hidden SU(2)$_H$ gauge symmetry. Once the SU(2)$_H$ gauge group is spontaneously broken down to U(1)$_H$ by an adjoint Higgs field, we have a 't Hooft-Polyakov monopole \cite{tHooft:1974kcl,Polyakov:1974ek} with a mass of order $M_M\simeq4\pi v/e_H$, as well as a heavy $W$ boson with a mass of $m_W=e_Hv\simeq\alpha_H M_M$, where $e_H$ is the gauge coupling for U(1)$_H$, $\alpha_H\equiv e_H^2/4\pi$, and $v$ is the VEV of the adjoint Higgs. We also assume that the axion couples to the SU(2)$_H$ as well as the U(1)$_H$ gauge field via anomalous couplings.

We assume that 
the SU(2)$_H$ symmetry is spontaneously broken after inflation, and monopoles (and anti-monopoles) are produced by the Kibble-Zurek mechanism~\cite{Kibble:1976sj,Zurek:1985qw}. In this paper, we do not go into the details of the monopole production, which has been discussed in detail in Refs.~\cite{Baek:2013dwa,Khoze:2014woa}. 
The axion we consider is not the QCD axion, but an axion-like particle with a constant mass in vacuum.
The axion starts to roll down to the potential minimum when the Hubble parameter decreases to its mass. 
We assume that this happens in the presence of the monopoles in background. 
In this case, the dissipation effect on the axion via the Schwinger effect can be relevant during the axion oscillation, and modifies the resulting axion abundance, as we will discuss in this paper.

%%%%%%%%%%%%%%%%%%%%%%%%%%%%%%%%%%%%%%%%%%%%%%%%%%%%%%%%%%%
%%%%%%%%%%%%%%%%%%%%%%%%%%%%%%%%%%%%%%%%%%%%%%%%%%%%%%%%%%%
\subsection{Witten effect
\label{sec:witten}}
In this subsection, we review the Witten effect on the axion in the monopole background. 
The Lagrangian with a CP violating $\Theta$-term for a hidden U$(1)_H$ gauge theory is given by 
\beq
\Lag\supset-\frac{1}{4}F_{\mu\nu}F^{\mu\nu}-\frac{e_H^2\Theta}{32\pi^2}F_{\mu\nu}\tilde{F}^{\mu\nu},
\eeq
where 
$F_{\mu\nu}$ and $\tilde{F}^{\mu\nu}\equiv\epsilon^{\mu\nu\lambda\rho}F_{\lambda\rho}/2$ represent the field strength of the hidden gauge field and its dual, respectively. 
The $\Theta$-term in an Abelian gauge theory is usually discarded since it is just a total derivative term and there is no non-trivial vacuum structure. However, it has a physical effect that becomes apparent in the presence of monopoles.  The $\Theta$-term induces an electric charge of the monopole and dyons as $Q_E/e_H=k-\Theta/2\pi$, where $k$ is an integer. Thus the monopole 
becomes dyon when $k - \Theta/(2\pi) \ne 0$~\cite{Witten:1979ey}.

Including the axion that couples to U(1)$_{H}$, the $\Theta$-term as well as the Lagrangian for the axion is given by 
\beq
\Lag_\phi= - \frac{1}{2} \del_\mu \phi \del^\mu \phi -V_\phi(\phi) -\frac{e_H^2}{32\pi^2}\frac{\phi+\Theta f_H}{f_H}F_{\mu\nu}\tilde{F}^{\mu\nu},
\eeq
where $f_H$ is the axion decay constant representing the strength of the above interaction. 
We assume that the axion has a potential of 
\beq
V_\phi(\phi)=m_\phi^2 f_\phi^2\left[ 1-\cos\lmk\frac{\phi}{f_\phi}\rmk \right],
\label{V_phi}
\eeq
where $m_\phi$ is the axion mass and $f_\phi$ is the axion decay constant for the potential.
We have introduced 
two different decay constants, $f_H$ and $f_\phi$. In the following we assume that there is a large hierarchy, $f_H \ll f_\phi$.\footnote{
\label{footnote:alpha} 
If the axion potential is mainly generated by the SU$(2)_H$ instanton effect, we would obtain $f_\phi \sim f_H$. 
In this paper, we conservatively assume $\alpha_H\lesssim0.1$, so that this contribution 
is exponentially suppressed and is negligible~\cite{Fuentes-Martin:2019bue,Csaki:2019vte,Buen-Abad:2019uoc}.
}
This can be realized by the clockwork mechanism~\cite{Kim:2004rp,Choi:2014rja,Higaki:2014qua,Higaki:2015jag,Choi:2015fiu,Kaplan:2015fuy,Giudice:2016yja,Higaki:2016yqk,Farina:2016tgd,Long:2018nsl,Chiang:2020aui}.\footnote{
It was first pointed out in Ref.~\cite{Higaki:2016yqk} that
the interaction between the hidden U(1) gauge field and the axion can be enhanced
in the clockwork mechanism.}

In the presence of the axion, the electric charge of the dyon becomes field-dependent as
\beq
\frac{Q_E}{e_H}=k-\frac{\phi+\Theta f_H}{2\pi f_H}.
\label{charge}
\eeq
As the value of the axion field changes, so does the electric charge of the dyon. Including the energy of the static electric field, the mass (or the total energy) of the dyon is given by
\beq
M_D (Q_E) &\simeq&
M_M+\frac{Q_E^2}{8\pi}m_W\label{dyonmass}
\\
&=&M_M+\frac{e_H^2}{8\pi}m_W\left(k-\frac{\phi+\Theta f_H}{2\pi f_H}\right)^2.
\eeq
If the number density of dyons is sufficiently high, the system can be spatially averaged. Then the electrostatic field energy density can be interpreted as the effective potential of the axion~\cite{Fischler:1983sc},
\beq
V_M(\phi) 
&=& \frac{1}{8} \alpha_H^2 \rho_M \lmk\frac{Q_E}{e_H}\rmk^2 \nonumber
\\
&=& \frac{1}{2}m_{\phi, M}^2(\phi +\Theta f_H -2\pi kf_H)^2
\label{effpot}
\eeq
with
\beq
m^2_{\phi, M}=\frac{\alpha_H^2}{16\pi^2}\frac{\rho_M}{f_H^2},
\eeq
where $\rho_M(T)$ is the energy density of the monopole and the anti-monopole at temperature $T$ of the SM sector, and we have assumed that $k$ is the same for all the monopoles.
The static electric field energy density depends on the mode $k$ for the dyon state. 
The potential energy \eq{effpot} is minimized when the dyon charge is the smallest. 
We therefore assume that $k$ is chosen such that the electric charge is minimized when a monopole is generated in the phase transition.\footnote{Strictly speaking, we need to determine how the mode $k$ is chosen  by numerical simulation of the  phase transition. In general, $k$ can take on multiple values. 
We expect that the dynamics of the phase transition proceeds in such a way as to minimize the energy of the entire system and assume that most monopoles have the same $k$ value that minimizes the energy.}

Including the effective potential \eq{effpot}, the total potential for the axion is given by
\beq
\label{eq:axionV}
V(\phi)=
m_\phi^2 f_\phi^2\left[1-\cos\lmk\frac{\phi}{f_\phi}\rmk\right] + \frac{1}{2}m_{\phi, M}^2(\phi +\Theta f_H -2\pi kf_H)^2.
\eeq
In the early universe, when the second term dominates over the first one, the axion dynamics becomes non-trivial. This was studied in Refs.~\cite{Kawasaki:2015lpf,Nomura:2015xil,Kawasaki:2017xwt,Nakagawa:2020zjr}, where it was found that the axion abundance can be suppressed due to the earlier onset of the oscillation and the adiabatic suppression mechanism.
The analyses in these papers can be justified in the case that the electric charge of the dyon never exceeds the Schwinger pair production threshold.
However, this is not the case if, for instance, $f_\phi \gg f_H$, as we will see shortly. 
In the following, we will study this complementary scenario with $f_\phi \gg f_H$.

%%%%%%%%%%%%%%%%%%%%%%%%%%%%%%%%%%%%%%%%%%%%%%%%%%%%%%%%%%%
%%%%%%%%%%%%%%%%%%%%%%%%%%%%%%%%%%%%%%%%%%%%%%%%%%%%%%%%%%%
\subsection{Dyon evaporation via the Schwinger effect
\label{sec:dyon}}
Here we discuss what happens if the axion field value changes coherently with its motion by the first term of \eq{eq:axionV}, while neglecting the effect of $V_M(\phi)$ on the axion dynamics for the moment.
The detailed dynamics of axion will be discussed in Sec.~\ref{sec:dynamics0}.

The axion motion changes the electric charge of the dyon.  
When the electrostatic energy for a dyon exceeds the $W$ boson mass, it is energetically preferred for the dyon with charge $|Q_E|$ to evaporate into a dyon with charge $|Q_E-e_H|$ and a $W$ boson with charge $e_H$.\footnote{This process is similar to black hole evaporation via the Hawking radiation \cite{Hawking:1974rv,Hawking:1975vcx}, and this is why we call it evaporation.}
The evaporation can be understood as the Schwinger pair production of $W^+ W^-$ bosons near the dyon surface and the absorption of a single $W$ boson (that has the opposite charge with the dyon) into the dyon. 
Noting that the source of the energy originates from the axion coherent motion, the dissipative effect on the axion dynamics is expected to be induced as the backreaction.

Let us consider the condition that the Schwinger process happens near the surface of a dyon, where 
a dyon with charge $Q_E = \pm ne_H$ evaporates into a dyon with charge $\pm(n-1)e_H$ and a $W^\pm$ boson. 
This process is energetically allowed if the following condition is met: 
\beq
 M_D(ne_H) - M_D((n-1)e_H) > m_W, 
\eeq
where we have assumed that the final-state particles are separated enough with each other and neglected the binding energy. 
We approximate the condition by assuming $n \gg 1$ and using \eq{dyonmass} and obtain 
\beq
\label{cond for Schw}
 \left|\frac{dM_D}{dQ_E}\right|
\gtrsim \frac{m_W}{e_H}  ~~ \Leftrightarrow ~~ |Q_E/e_H| \gtrsim \alpha_H^{-1}.
\eeq
Here let us mention a caveat in this analysis. The above calculations were based on the assumption that the internal structure of the 't Hooft Polyakov monopole does not change as $Q_E$ increases. This assumption, however, becomes incorrect as $Q_E$ exceeds the threshold value.  For $e_H |Q_E| > 4\pi$, the radius of the monopole is expected to increase in proportion to $\sqrt{Q_E}$. This is because the core radius is determined in such a way that the energy of the core balances the energy of the electromagnetic field outside the core. As a result, the left-hand side of \eq{cond for Schw} increases more slowly for  $Q_E$ above the threshold. In any case, the pair production of $W$ bosons due to the Schwinger effect should occur for $Q_E$ not very different from the critical value.\footnote{When the dyon remains stable against decay for $Q_E$ above the threshold value, its core radius and mass will continue to increase as the axion field moves, and the axion potential will also increase in proportion to $\phi^{7/2}$.
We will discuss elsewhere when this case has a significant impact on axion dynamics.
} In what follows, we assume that the creation of $W$ boson pairs occurs when the condition (\ref{cond for Schw}) is met, but keep in mind that there is an uncertainty of order unity.
See Refs.~\cite{Julia:1975ff,Prasad:1975kr} for numerical and analytic studies of dyon solution and its energy.

As we discussed in Sec.~\ref{sec:witten}, 
the value of $k$ in \eq{effpot} is fixed at the formation of monopoles such that $V_M(\phi)$ is minimized for fixed $\phi$ and $\Theta$. 
Thus, the electric charge of the monopole should have $|Q_E / e_H| < 1/2$ 
before the axion starts to oscillate. 
This also means that the axion field value is close to (but not exactly at) the minimum of the potential induced by the Witten effect. 
As the axion begins to oscillate by the first term of \eq{eq:axionV}, the electric charge of the dyon increases (see \eq{charge}) and eventually crosses the threshold of the Schwinger process given by \eq{cond for Schw}. Then, the dyon starts to evaporate when the axion field value changes by the amount
\beq
\delta\phi_{\rm first}\equiv2\pi \alpha_H^{-1} f_H. 
\label{shift}
\eeq
Our interest lies in the scenario where this happens many times during the axion oscillation.
Since the oscillation is driven by the potential given in \eq{V_phi},
its typical amplitude 
is of the order of $f_\phi$. We therefore consider the case of 
\beq
\frac{f_\phi}{f_H}\gg
\frac{2\pi}{\alpha_H}.
\label{condition}
\eeq
For $\alpha_H \lesssim 0.1$ (see footnote~\ref{footnote:alpha}), the above requires $f_\phi/f_H \gg 10^{2}$. 
This substantial hierarchy can be realized by the clockwork mechanism~\cite{Kim:2004rp,Choi:2014rja,Higaki:2014qua,Higaki:2015jag,Choi:2015fiu,Kaplan:2015fuy,Giudice:2016yja,Higaki:2016yqk,Farina:2016tgd,Long:2018nsl,Chiang:2020aui}.

When the axion field value changes by the amount $\delta\phi_{\rm first}$, each dyon emits a single $W$ boson, thereby reducing its electric charge by a single unit. This process provides a backreaction to the axion dynamics, which we will explore in Sec.~\ref{sec:rate}. For the moment, let us ignore the back reaction to the axion dynamics and assume that the axion continues to roll down to the potential minimum of the first term in \eq{eq:axionV}. After a further shift of the axion field value by 
\beq 
 \delta\phi=2\pi f_H, 
\eeq
the electric charge of the dyons reaches the threshold for the Schwinger process again. Each dyon then emits another single $W$ boson, reducing its electric charge by a single unit. This process continues until the axion reaches the endpoint and then begins to move in the opposite direction.

\begin{figure}[t!]
\includegraphics[width=11cm]{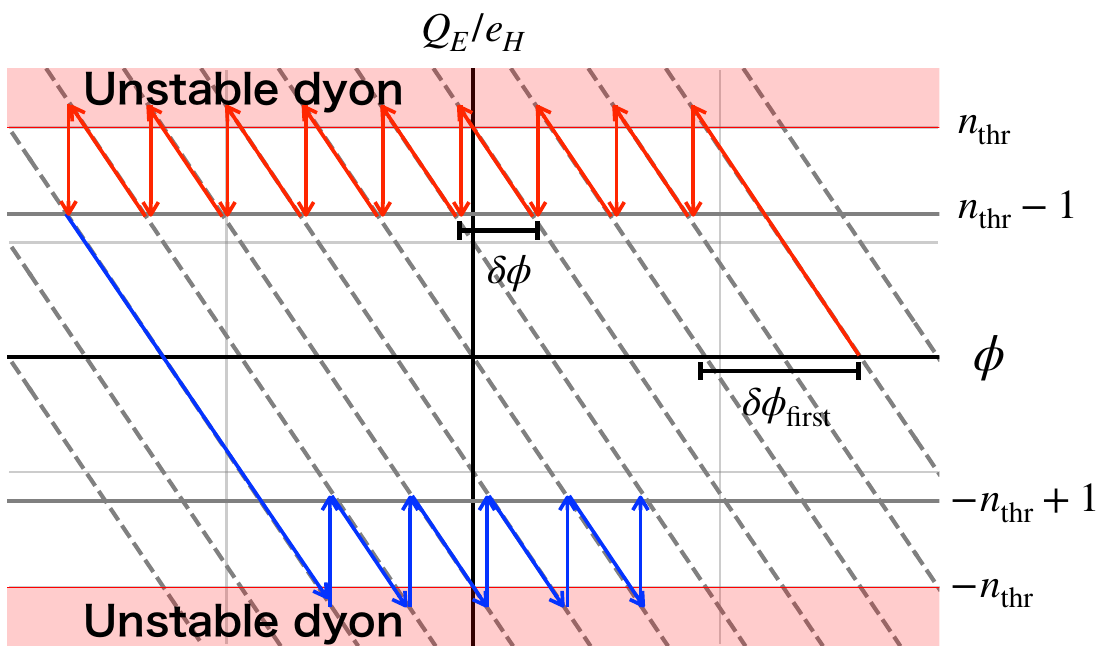}
\centering
\caption{
A schematic picture for the evolution in $\phi$-$Q_E/e_H$ plane by the dyon evaporation. The gray dashed lines represent the allowed dyon states. In the red shaded region (that is given by $|Q_E/e_H| \geq n_{\rm thr}$), the dyon becomes unstable and emits a $W$ boson. The red and blue arrows represent the transition of the dyon state as the axion oscillates.
}
\label{fig:dyon}
\end{figure}

This sequence of events is shown schematically as the red arrows in \FIG{fig:dyon}. This figure illustrates the evolution of the electric charge of the dyon, $Q_E$, and the axion field value, $\phi$, during the first half cycle of the axion oscillation. The gray dashed lines represent the allowed dyon states according to \eq{charge}. The dyon is unstable against evaporation in the red shaded regions, $|Q_E|/e_H \geq n_{\rm thr}$, where $n_{\rm thr}$ denotes the threshold dyon mode close to $\alpha_H^{-1}$. 
Note that the scale in this figure is exaggerated for clarity. In particular, the gray dashed lines should appear much denser in a realistic case.

Since the axion oscillates around the minimum of its potential \eq{V_phi}, it begins to move in the opposite direction after the first half cycle.  Then the electric charge of the dyons begins to decrease in size. 
Eventually, the electric charge crosses zero and increases in size again with the opposite sign. When the axion field value changes by a factor of $2\delta\phi_{\rm first}$ in the second half cycle, the electric charge of the dyon reaches the threshold of the Schwinger process and the evaporation takes place again.  These processes are represented by the blue arrows in \FIG{fig:dyon}.  These processes are repeated during the axion oscillation, whose amplitude decreases with time due to the cosmological redshift and the dissipation effect we will discuss shortly.  When the amplitude of the axion oscillation becomes smaller than $2\delta\phi_{\rm first}$, the evaporation of the dyons stops.

\subsection{Absorption 
and annihilation 
of $W$ bosons}
\label{sec:Wboson}
In the previous subsection we have seen that the dyon evaporation proceeds via the production of $W$ bosons by the Schwinger effect. 
In this subsection, we consider the annihilation process of $W^\pm$ bosons and the inverse process of the dyon evaporation which is the absorption of $W$ bosons by dyons. 
We will see shortly that this inverse process is negligible. 
One can skip this subsection if this conclusion is admitted.

We focus on the dynamics of the $W^\pm$ boson after their production. 
First, we consider the interaction between a $W^\pm$ boson and a dyon (or an anti-dyon) that produces the $W^\pm$ boson via the evaporation. 
Since the $W^\pm$ boson has the electric charge and the magnetic dipole moment, 
it feels both electric and magnetic potentials of the dyon (anti-dyon). 
The magnetic interaction results in  
an attractive force with the potential $U(r)=-1/(2m_Wr^2)$. 
This is weaker than the electric Coulomb force and is negligible outside the dyon (i.e., for $r > m_W^{-1}$)
for the monopole with charge $|Q_E|/e_H\sim\alpha_H^{-1}$. 
Since the electric potential near the surface of the dyon (anti-dyon) is as large as the $W^\pm$ boson mass, 
the emitted $W^\pm$ boson 
is accelerated 
to a high speed, close to the speed of light. 
Their dynamics is driven mainly by the Coulomb force.

Next, we consider the dynamics of the $W^\pm$ bosons in the dyon and anti-dyon background. For that purpose, let us estimate the typical separation length between dyon and anti-dyons. 
It can be roughly estimated by the number density of the dyon. 
At the onset of the axion oscillation, $T=T_{\rm osc}$, the number density of the monopole is given by
\beq
n_M(T_{\rm osc})=\frac{\rho_{M,0}}{M_M}\frac{s_{\rm osc}}{s_0},
\eeq
where $s_{\rm osc}$ and $s_0$ denote the entropy density at the onset of the axion oscillation and at present, respectively, and $\rho_{M,0}$ is the monopole energy density at present.
We assume that the universe is radiation dominated and the onset of the axion oscillation is roughly given by $m_\phi \sim H$, where $H=\dot{a}/a$ is defined as the Hubble parameter and $a(t)$ as the scale factor. 
The separation length at $T = T_{\rm osc}$ can then be estimated as 
\beq
d&\simeq& n_M^{-1/3}(T_{\rm osc})=\left(\frac{g_{*s0}}{g_{*s}(T_{\rm osc})}\frac{M_M}{\Omega_M\rho_{\rm crit}}\right)^{1/3}\frac{T_0}{T_{\rm osc}}\nonumber\\
&\simeq& 3.0 \times 10^3 \GeV^{-1}\left(\frac{g_{*s}(T_{\rm osc})}{106.75}\right)^{-\frac{1}{12} }\left(\frac{\Omega_M}{10^{-3}\Omega_{\rm DM}}\right)^{-\frac{1}{3}}
\nonumber\\&&\times
\left(\frac{\alpha_H}{10^{-2}}\right)^{-\frac{1}{3}} \left(\frac{m_W}{100\GeV}\right)^{\frac{1}{3}}\left(\frac{m_\phi}{1\mu{\rm eV}}\right)^{-\frac{1}{2}},
\eeq
where $T_0$ is the temperature at present, and $\rho_{\rm crit}$ is the critical density. One can see that $d\ll m_\phi^{-1}$, which implies that produced $W^\pm$ bosons 
will meet other $W^\pm$ bosons and neighboring anti-dyon (dyon) within the time scale much shorter than the oscillation period of axion, $m_\phi^{-1}$. 
Therefore, the scattering processes (including annihilation and absorption) among $W^\pm$ bosons and dyons (anti-dyons) should be taken into consideration.

$W^{\pm}$ bosons in this setup can annihilate into dark photons. 
If $W^{\pm}$ bosons are produced abundantly via the evaporation of dyons,
the annihilation into dark photons becomes effective and the number density stops growing at a threshold given by $\sigma_{\rm ann} n_W = H$, where $\sigma_{\rm ann} \sim\alpha_H^2/m_W^2$ is the annihilation cross section.
This leads to an upper bound on the number density, 
\begin{equation}
\label{pairann}
n_W \lesssim \frac{m_\phi m_W^2}{\alpha_H^2},
\end{equation}
at $H \sim m_\phi$.%
\footnote{
The non-thermal production can cause the overproduction of $W$ boson.
However, when the freeze out follows the axion oscillation e.g. for $m_\phi\sim 1\mu{\rm eV}$ and $m_W\sim\mathcal{O}(100)\GeV$, the $W$ boson abundance is almost the same as that without extra production.
According to the analysis in \REF{Khoze:2014woa}, the $W$ boson abundance can be ignored for $\alpha_H\sim10^{-2}$ and $m_W\lesssim 100\GeV$.
Thus the $W$ boson abundance is negligibly small as long as $\alpha_H\sim10^{-2}$, $m_W\lesssim \mathcal{O}(100)\GeV$, and $m_\phi\sim1\mu{\rm eV}$.
}
Hereafter we consider the case where the $W$ boson abundance satisfies Eq.~(\ref{pairann}).

The absorption of the $W$ bosons into dyons occurs 
via the following processes: (i) $D_n + W^-\rightarrow D_{n-1} + \gamma'$ and (ii) $D_{n-1} + W^+ \rightarrow D_n + \gamma'$, 
where 
we denote a dyon with charge $ne_H$ as $D_n$ and dark photon as $\gamma'$.
According to \cite{Smilga:1988pk}, a non-relativistic $W$ boson is absorbed into
a dyon with a cross section of $\sigma_{\rm abs}\sim\alpha_H/m_W^2$. 
We assume this is also the case for our case, where $W^\pm$ bosons are marginally relativistic.

Now we shall compare the rates of $W$ boson production via the Schwinger process and that of the absorption process. 
Note that 
approximately $(\phi_{\rm amp}/f_H)$ $W$ bosons are produced as the axion moves from $+\phi_{\rm amp}$ to $-\phi_{\rm amp}$ within the half period, $\pi/m_\phi$. 
Then their rates per unit dyon are estimated as 
\begin{align}
\Gamma_{\rm prod} &\sim \frac{m_\phi}{\pi} \frac{2\phi_{\rm amp}}{f_H}, \\
\Gamma_{\rm abs} &\sim \frac{\alpha_H}{m_W^2} n_W,
\end{align}
at the onset of axion oscillation, $H \sim m_\phi$. 
Assuming Eq.~(\ref{pairann}) and $\phi_{\rm amp} \gg f_H$, we find that $\Gamma_{\rm prod} \gg \Gamma_{\rm abs}$. 
This implies that 
the absorption process (i.e., the inverse process) is negligible compared with the production process via evaporation. 
This ensures that the evaporation of dyons can be interpreted as the dissipation of axion energy into background dark photons.

\section{Effect on axion dynamics}
\label{sec:dynamics0}
In this section, we examine the effect of the dyon evaporation on the evolution of axion. In particular we numerically estimate its impact on
the axion abundance.

\subsection{Typical time scales for axion dynamics
\label{sec:mass}}

First, let us briefly consider the time evolution of the axion field in the absence of dyon evaporation.
At  high temperatures, we expect $V_M$ dominates over $V_\phi$. 
When the Hubble parameter becomes comparable to the mass from the Witten effect, the axion starts to oscillate around the minimum of $V_M$. 
The temperature at the onset of this oscillation, $T_M$, is estimated by solving $3H=m_{\phi,M}$:
\beq
T_M &=& \frac{5\alpha_H^2}{8\pi^4g_{*s0}}\frac{\rho_{M0}\Mpl^2}{f_H^2T_0^3}\nonumber\\
&\simeq& 71{\rm GeV} \lmk\frac{\alpha_H}{10^{-2}}\rmk^2 \lmk\frac{\Omega_M}{10^{-3}\Omega_{\rm DM}}\rmk \lmk\frac{f_H}{10^{8}\GeV}\rmk^{-2}.
\eeq
The initial amplitude is of the order of $f_H$, and the oscillation energy is diluted by the cosmic expansion.  In the following, we neglect the energy of these initial oscillations, since it is much smaller than the energy of later oscillations due to $V_\phi$.

Similarly, we define $T_{\rm osc}$ by $3H(T_{\rm osc})=m_\phi$. This corresponds to the timing when the axion starts to oscillate in the potential $V_\phi$. In the following we focus on the case of $T_M > T_{\rm osc}$, since otherwise
the potential $V_M$ rarely affects the dynamics of the axion
and the dissipation effect is also negligibly small.

As the universe expands, $V_\phi$ becomes dominate over $V_M$ and then the potential minimum begins to approach the origin, which is the minimum of $V_\phi$. The temperature at this moment, $T_{\rm shift}$, is estimated by $m_\phi=m_{\phi,M}$, which gives
\beq
T_{\rm shift} &=& \lmk\frac{16\pi^2g_{*s0}m_\phi^2f_H^2}{g_{*s}(T_{\rm shift})\alpha_H^2\rho_{M0}}\rmk^{1/3}T_0\nonumber\\
&\simeq& 10\GeV \lmk\frac{g_{*s}(T_{\rm shift})}{80}\rmk^{-1/3}\lmk\frac{\alpha_H}{10^{-2}}\rmk^{-2/3}\lmk\frac{\Omega_{M}}{10^{-3}\Omega_{\rm DM}}\rmk^{-1/3} \lmk\frac{m_\phi}{1\mu{\rm eV}}\rmk^{2/3} \lmk\frac{f_H}{10^8\GeV}\rmk^{2/3}.\nonumber\\
\label{Tshift}
\eeq
We note that the time scale of the potential change is of order $H^{-1}$, while the time scale of the axion oscillation is much faster at a time around $T = T_{\rm shift}$. 
This implies that the potential changes adiabatically. 
It is known that, if the evaporation of dyons is neglected, the axion number density produced by the adiabatic potential shift 
is strongly suppressed by an exponential factor~\cite{Linde:1996cx}. 
This mechanism has been considered for the axion model with the Witten effect in Refs.~\cite{Kawasaki:2015lpf,Kawasaki:2017xwt,Nakagawa:2020zjr}.

\begin{figure}[t!]
\includegraphics[width=11cm]{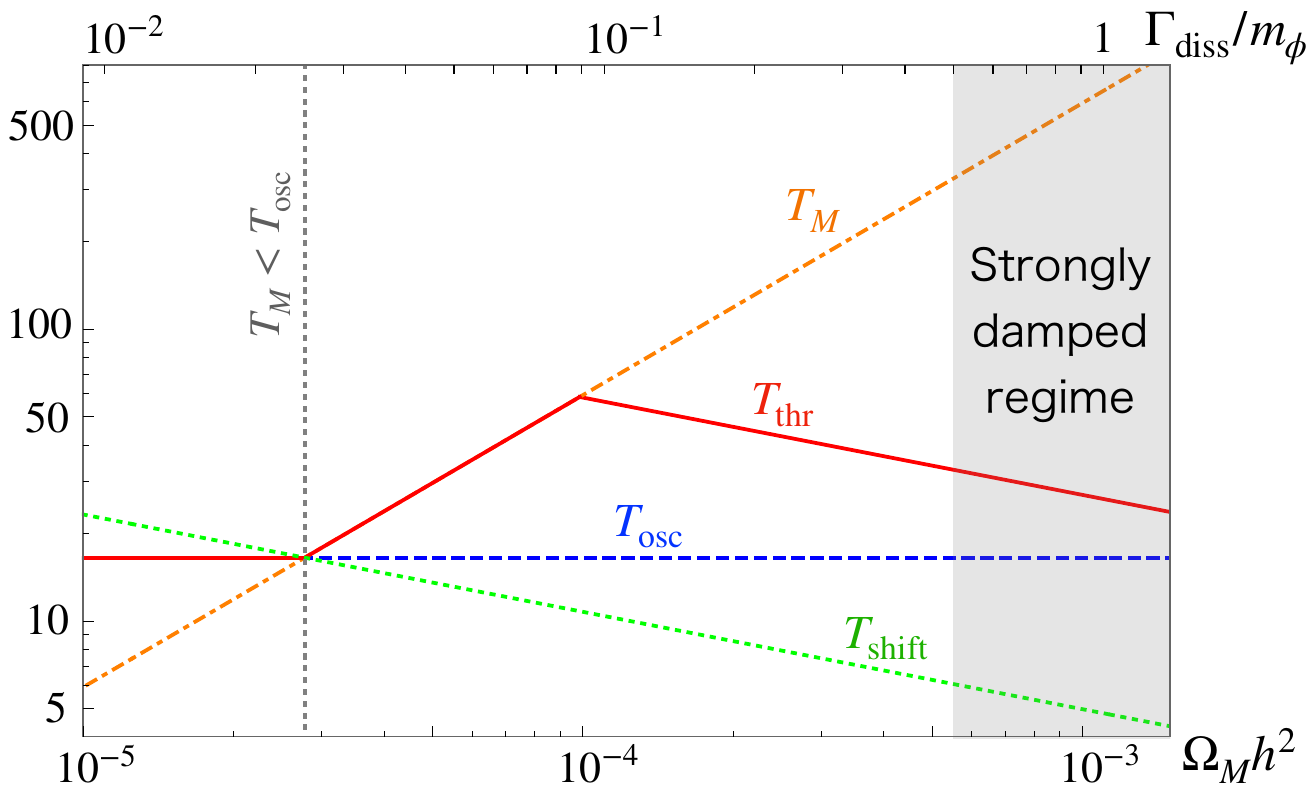}
\centering
\caption{
Various temperatures as a function of $\Omega_Mh^2$ [bottom] ($\Gamma_{\rm diss}(T_{\rm osc})/m_\phi$[top]).
We take $m_\phi=1\mu{\rm eV}$, $\phi_{\rm ini}=10^{13}\GeV$, $f_H=10^8\GeV$, and $\alpha_H=10^{-2}$.
The red solid, green dotted, blue dashed, and orange dot-dashed lines denote $T_{\rm thr}$, $T_{\rm shift}$, $T_{\rm osc}$, and $T_M$ in the unit of GeV, respectively.
The temperature at the end of the dissipation, $T_{\rm end}$, is about two orders of magnitude smaller than the temperature shown in this figure.
}
\label{fig:temp}
\end{figure}

Next we consider the evaporation of dyons.  
As the axion moves, the magnitude of the electric charge of the dyon increases.  When the axion moves over a field range, $\delta\phi_{\rm first}$, 
the dyon obtains a mass and charge over the instability threshold, and starts to evaporate.
Let us define this time as $T_{\rm thr}$.
If $V_M$ is large enough, 
$T_{\rm thr}$ is determined by the condition that the deviation from the initial position is equal to $\delta\phi_{\rm first}$:
\beq
T_{\rm thr} &=& T_0\lmk\frac{8\pi}{\alpha_H}\frac{g_{*s0}}{g_{*s}(T_{\rm thr})} \frac{m_\phi^2f_H\phi_{\rm ini}}{\rho_{M0}}\rmk^{1/3}~~~~(<T_M)\nonumber\\
&\simeq& 55\GeV \lmk\frac{g_{*s}(T_{\rm thr})}{80}\rmk^{-1/3} \lmk\frac{\alpha_H}{10^{-2}}\rmk^{-1/3} 
\lmk\frac{f_H}{10^8\GeV}\rmk^{1/3}
\lmk\frac{\Omega_M}{10^{-3}\Omega_{\rm DM}}\rmk^{-1/3}\nonumber\\
&& \times\lmk\frac{m_\phi}{1\mu{\rm eV}}\rmk^{2/3}
\lmk\frac{\phi_{\rm ini}}{10^{13}\GeV}\rmk^{1/3}.
\label{eq:T_{thr}}
\eeq
The typical values of the temperatures defined above are shown in \FIG{fig:temp} as a function of the monopole abundance. We take
$m_\phi=1\mu{\rm eV}$, $\phi_{\rm ini}=10^{13}\GeV$, $f_H=10^8\GeV$, and $\alpha_H=10^{-2}$.
The red solid, green dotted, blue dashed, and orange dot-dashed lines denote $T_{\rm thr}$, $T_{\rm shift}$, $T_{\rm osc}$, and $T_M$ in the unit of GeV, respectively.
The vertical dotted line represents $T_M = T_{\rm osc}$. 
The red line at $\Omega_M h^2 \gtrsim 10^{-4}$ corresponds to \eq{eq:T_{thr}}. 
For smaller $\Omega_M h^2$, 
$V_M$ is too small to satisfy $T_{\rm thr} < T_M$. 
In that case, $T_{\rm thr} \simeq {\rm Max}[T_M, T_{\rm osc}]$.

On the upper horizontal axis, we show $\Gamma_{\rm diss}(T_{\rm osc})/m_\phi$, where $\Gamma_{\rm diss}$ is the dissipation rate to be defined shortly. 
As we will see in \SEC{sec:evolution}, the dissipation effect is too weak to affect the axion dynamics for $T_M\ll T_{\rm osc}$. 
The gray shaded region is the regime where the expression of $\Gamma_{\rm diss}$ is not applicable. This will be explained in Sec.~\ref{sec:rate} in more detail. We are interested in the range where the dissipation rate is moderately large, $\Gamma_{\rm diss}/m_\phi\sim\mathcal{O}(0.1)$.

We note that the temperature $T_{\rm shift}$ is close to $T_{\rm osc}$ and $T_{\rm thr}$ in the parameter region of our interest. This implies that $V_\phi$ and $V_M$ are comparable in size to each other when the evaporation occurs. 
Thus we can roughly approximate that the time scale of the axion motion is determined by the potential $V_\phi$ (omitting $V_M$) for simplicity.\footnote{
In addition, as mentioned earlier, it is possible that the actual threshold of the dyon instability is slightly larger than $Q_E/e_H=\alpha_H^{-1}$, which could delay the first evaporation to take place, reducing $T_{\rm thr}$. In this case our approximation 
is better.
}
This approximation will be used in our numerical calculation in Sec.~\ref{sec:dynamics}.

\subsection{Cascade process for evaporation
\label{sec:cascade}}

As explained in \SEC{sec:dyon}, the electric charge of dyons is reduced by $e_H$ when the first evaporation takes place. 
Then the effective potential from the Witten effect, $V_M$, suddenly changes as
\beq
V_M(\phi) 
&=& \frac{1}{2}m_{\phi, M}^2(\phi +\Theta f_H -2\pi kf_H)^2
\\
&\to& \frac{1}{2}m_{\phi, M}^2(\phi +\Theta f_H -2\pi (k - 1) f_H)^2, 
\eeq
where we have taken
$Q_E/e_H > 0$ at the first step of evaporation without loss of generality. 
Thus the axion feels a sudden change of the potential height, 
and the minimum of the total axion potential is shifted by a factor of $m_{\phi,M}^2\delta\phi/(m_{\phi,M}^2+m_\phi^2)$ toward the origin. 
We assume that this process happens much faster than the time scale of axion coherent motion. 
Then the axion field value does not change while its potential changes. 
This implies that the axion starts to oscillate around the new potential minimum with the amplitude of the same amount: $m_{\phi,M}^2\delta\phi/(m_{\phi,M}^2+m_\phi^2)$.

If $m_{\phi,M}^2/(m_{\phi,M}^2+m_\phi^2) > 1/2$, 
the axion moves by a factor of more than $\delta \phi$ within the half period of its coherent oscillation.
This means that the subsequent evaporation occurs 
within the half period, 
and the axion potential $V_M$ again changes 
and induces an extra amplitude for the axion coherent motion. 
At this second evaporation, the axion has the extra amplitude (or potential energy) in addition to the kinetic energy, which are enough to induce subsequent evaporation. 
These processes continue to occur as a chain reaction. After the axion reaches the other endpoint, it starts to move in the opposite direction. Then, the similar dissipation process continues until the amplitude becomes smaller than $2 \delta \phi_{\rm first}$. See Fig.~\ref{fig:dyon}.

Here we note that the adiabatic condition is violated at every step for evaporation because of the sudden change of the effective potential. 
One could say that, 
compared with the case without evaporation due to the Schwinger process, 
the axion abundance is {\it enhanced}. 
Still, one has to take into account another (opposite) effect of evaporation on the axion dynamics, which we will discuss in Sec.~\ref{sec:rate}.

\subsection{Dissipation rate
\label{sec:rate}}
Now we estimate how strongly the evaporation process drives axion energy dissipation as a backreaction.
When the axion  shifts by $\Delta\phi~(>\delta \phi)$, the change of the energy density is estimated as 
\beq
\Delta\rho_\phi = -n_Mm_W\cdot\frac{\Delta\phi}{\delta\phi},
\label{eq:deltarho}
\eeq
because of the conservation of total energy. 
Here we have assumed that the energy density of the axion is greater than that of the $W$ boson produced in a single evaporation, otherwise the dyon evaporation cannot proceed.
This condition requires $\rho_\phi/\rho_M>\alpha_H$ at the beginning of the axion oscillation.
Dividing both sides by $\Delta t$ and taking the limit of $\Delta t\rightarrow0$, we obtain the dissipation rate of the axion energy density as a backreaction of the dyon evaporation as
\beq
\left.\frac{d\rho_\phi}{dt}\right|_{\rm diss}=-\frac{n_Mm_W}{\delta\phi}|\dot{\phi}|.
\eeq
If \eq{condition} is satisfied, 
the axion field value changes by a factor much larger than $\delta \phi$ within a period of oscillation, and 
this continuous limit is a good approximation. 
The time evolution of the energy density is therefore written as 
\beq
\dot{\rho}_\phi+3H \rho_\phi =-\frac{n_Mm_W}{\delta\phi}|\dot{\phi}|.
\label{conserve}
\eeq
The right-hand side represents the dissipation of the axion energy by the backreaction of the dyon evaporation.

To solve Eq.~(\ref{conserve}) numerically or analytically, 
it is convenient to adopt approximations by assuming that the evolution of $\phi$ is determined mainly by the mass term $m_\phi^2 \phi^2/2$.
After taking time average over the oscillation period, 
we approximate $|\dot{\phi}| \simeq \sqrt{\rho_\phi}$ by using the virial theorem.
Then we can rewrite the equation as 
\beq
\dot{\rho}_\phi+3H \rho_\phi  \simeq -\frac{n_Mm_W}{\delta\phi}\sqrt{\rho_\phi}.
\label{approx}
\eeq
This approximated relation is useful to directly calculate the evolution of energy density of axion.

On the other hand, to examine the evolution of the axion field,
it is convenient to approximate the dissipation term  as 
\beq
-\frac{n_Mm_W}{\delta\phi}|\dot{\phi}| \simeq -\frac{n_Mm_W}{\delta\phi}\frac{\dot{\phi}^2}{m_\phi\phi_{\rm amp}}\equiv -\Gamma_{\rm diss}(T)\dot{\phi}^2,
\label{dissipation}
\eeq
where $\phi_{\rm amp}$ is the oscillation amplitude of axion. In the first equality, we again take time average of $|\dot{\phi}|$ over the oscillation period.
This approximation reduces the dissipation term to the canonical form of frictional force for the scalar field. 
Substituting the explicit form of $\rho_\phi$ and $P_\phi$ into \eq{conserve}, we obtain the equation of motion for the axion 
\beq
\ddot{\phi}+(3H+\Gamma_{\rm diss})\dot{\phi} + V_\phi'(\phi) = 0.
\label{eom}
\eeq
This form makes it obvious that the dyon evaporation results in the dissipation of the axion energy. 
We note that $\Gamma_{\rm diss}$ ($\propto a^{-3/2}$) decreases slower than $H$ ($\propto a^{-2}$) 
after the axion oscillates 
because $n_M \propto a^{-3}$ and $\phi_{\rm amp}^{-1} \propto a^{3/2}$. 
Therefore, it can dominate over the Hubble friction term at a later epoch. 
This will be explicitly demonstrated in \SEC{sec:dynamics}.

We note that \eq{eq:deltarho} can be rewritten in terms of $\Gamma_{\rm diss}$ as 
\beq
\frac{\Delta\rho_\phi}{m_\phi^2 \phi_{\rm amp}^2/2} = -  \frac{ 2 \Delta\phi }{\phi_{\rm amp}} \frac{\Gamma_{\rm diss}}{m_\phi},
\label{eq:deltarho2}
\eeq
where $m_\phi^2 \phi_{\rm amp}^2/2$ is the oscillation energy of the axion. 
Considering the first quarter period of the axion oscillation, we take $\Delta \phi \simeq \phi_{\rm amp}$. 
Then the dissipated energy is comparable to or exceeds the oscillation energy for $\Gamma_{\rm diss}/m_\phi \gtrsim 1/2$. 
In this case, the dissipation is too strong for the axion to oscillate around the origin of the potential. 
Since the axion does not oscillate and moves slowly toward the origin, we cannot justify the approximation basd on the virial theorem used above.
In this regime, we need a further detailed study, which we leave for a future work.
This regime is shown as the gray shaded region in Fig.~\ref{fig:temp}.

\subsection{End of evaporation} 

The dissipation process does not last forever. 
Due to the Hubble expansion and the dissipation effect, the axion amplitude decreases with time. 
Since the dyon evaporates when the axion moves over a field range larger than $2\delta\phi_{\rm first}$ during the half period of its oscillation, the dissipation ends when the oscillation energy becomes comparable to $m_\phi^2\delta\phi_{\rm first}^2/2$.  
We denote the temperature at this moment as $T_{\rm end}$, which is determined by 
\beq
\rho_\phi(T_{\rm end}) \simeq \frac{1}{2}m_\phi^2\delta\phi_{\rm first}^2. 
\label{endpoint}
\eeq
We can derive the lower bound on $T_{\rm end}$ by considering the case without the Witten effect and the dissipation due to the Schwinger effect.
In this case the oscillation amplitude decreases in proportion to $T^{-3/2}$,
and so, we have
\beq
\label{Tendlow}
T_{\rm end} \geq T_{\rm osc} \left(\frac{\delta \phi_{\rm first}}{\phi_{\rm ini}}\right)^{2/3}.
\eeq
We must solve the axion dynamics numerically to determine the exact value of $T_{\rm end}$. 
In the weakly damped regime, the oscillation amplitude gradually decreases to a value less than $\delta \phi_{\rm first}$, stopping dissipation.
On the other hand, there may be a case where the system later enters the strongly damped regime, and the amplitude soon becomes less than $\delta \phi_{\rm first}$, and dissipation stops. In both cases, after the dissipation stops, the axion oscillates around the potential minimum without any additional frictional force.

Another condition for the dissipation to occur is that the 
axion energy density should be larger than the energy density of $W$ boson produced per a single evaporation. In fact,  this condition is weaker than the condition, $T > T_{\rm end}$,
and so we ignore this condition in the following.

%%%%%%%%%%%%%%%%%%%%%%%%%%%%%%%%%%%%%%%%%%%%%%%%%%%%%%%%%%%
%%%%%%%%%%%%%%%%%%%%%%%%%%%%%%%%%%%%%%%%%%%%%%%%%%%%%%%%%%%
\section{Numerical calculations
\label{sec:evolution}}
In this section we study the axion dynamics in the presence of the dissipation term and estimate a resultant suppression factor for the axion abundance.  As we commented in Sec.~\ref{sec:mass}, we have found that the difference between $T_{\rm thr}$, $T_{\rm osc}$ and $T_{\rm shift}$ is small and they are comparable to each other. We can then consider that the axion oscillates through the potential of $V_\phi$ at $H \simeq m_\phi$. In particular, we neglect $V_M$ for simplicity, while including the dissipation term $-\Gamma_{\rm diss} \dot{\phi}$. Neglecting $V_M$ is also supported by the fact that the adiabatic condition is violated by the sudden change of the effective potential at each step of evaporation (see the discussion at the end of Sec.~\ref{sec:cascade}).

%%%%%%%%%%%%%%%%%%%%%%%%%%%%%%%%%%%%%%%%%%%%%%%%%%%%%%%%%%
%%%%%%%%%%%%%%%%%%%%%%%%%%%%%%%%%%%%%%%%%%%%%%%%%%%%%%%%%%
\subsection{Numerical simulation of axion dynamics
\label{sec:dynamics}}

\begin{figure}[t!]
\includegraphics[width=7.5cm]{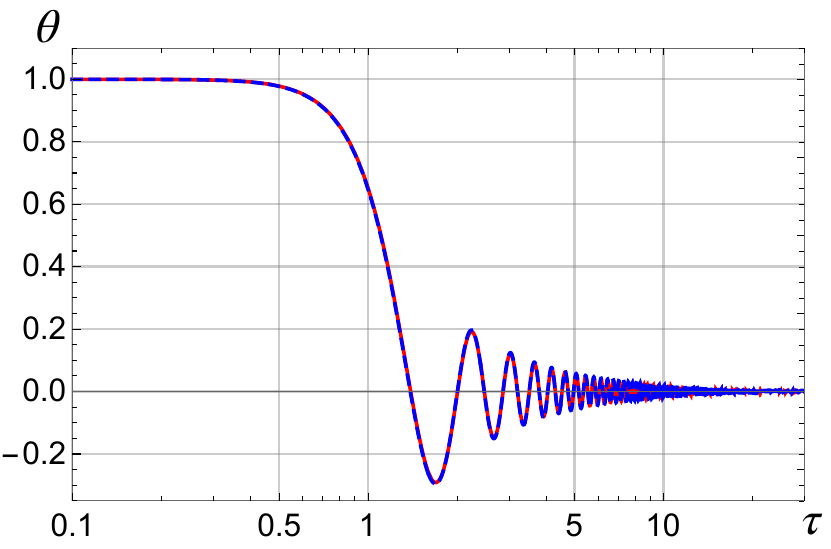}
\centering
\includegraphics[width=7.5cm]{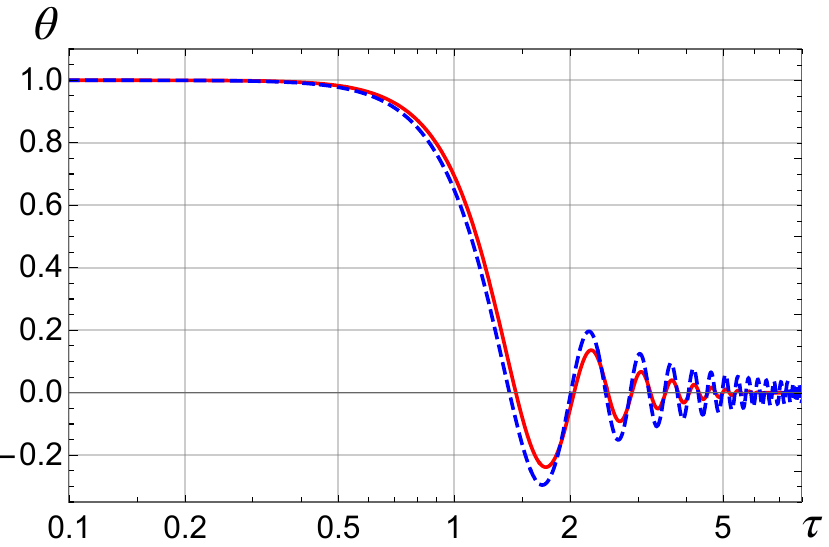}
\centering
\caption{Time evolution of the axion field as a function of $\tau$ ($\equiv T_{\rm osc}/T$) with (red solid) or without (blue dotted) dissipation.
We take $m_\phi=1\mu{\rm eV}$, $f_\phi=10^{13}\GeV$, $f_H=10^8\GeV$, $\alpha_H=10^{-2}$, and $\theta_{\rm ini}=1$. 
The strength of the dissipation effect is set to $\Gamma_{\rm diss}(T_{\rm osc})/m_\phi = 0.01$ and $0.15$ in the left and right panel, respectively.
In the left panel, the two lines overlap.
}
\label{fig:dynamics}
\end{figure}

We numerically solve the equation of motion (\ref{eom}) to obtain the time evolution of axion field $\theta \equiv \phi/f_\phi$. 
The result is shown in \FIG{fig:dynamics} as a function of an inverse temperature $\tau ~\equiv T_{\rm osc}/T$. 
The red solid (blue dashed) line represents the result with (without) dissipation. 
We take $m_\phi=1\mu{\rm eV}$, $f_\phi=10^{13}\GeV$, $f_H=10^8\GeV$, $\alpha_H=10^{-2}$, and $\theta_{\rm ini}=1$.
We use the fitting formulas for $g_*(T)$ and $g_{*s}(T)$ in \cite{Saikawa:2018rcs}.
In the left panel, we take $\Gamma_{\rm diss}(T_{\rm osc})/m_\phi = 0.01$ at the onset of the axion oscillation, where the dissipation does not significantly affect the axion dynamics.
In the right panel, we take $\Gamma_{\rm diss}(T_{\rm osc})/m_\phi = 0.15$. In this case, we can see that the oscillation amplitude is suppressed due to the dissipation. Note that the dissipation rate (\ref{dissipation}) depends on the parameters, $\alpha_H$, $\Omega_M$, $\theta_{\rm ini}$, $m_\phi$, $f_\phi$, and $f_H$. Here we vary $\Omega_M$ to change $\Gamma_{\rm diss}(T_{\rm osc})/m_\phi$. The corresponding values of $\Omega_M h^2$ are $1.1\times10^{-5}$ and $1.6\times10^{-4}$ for $\Gamma_{\rm diss}(T_{\rm osc})/m_\phi =0.01$ and $0.15$, respectively.

\begin{figure}[t!]
\includegraphics[width=7.7cm]{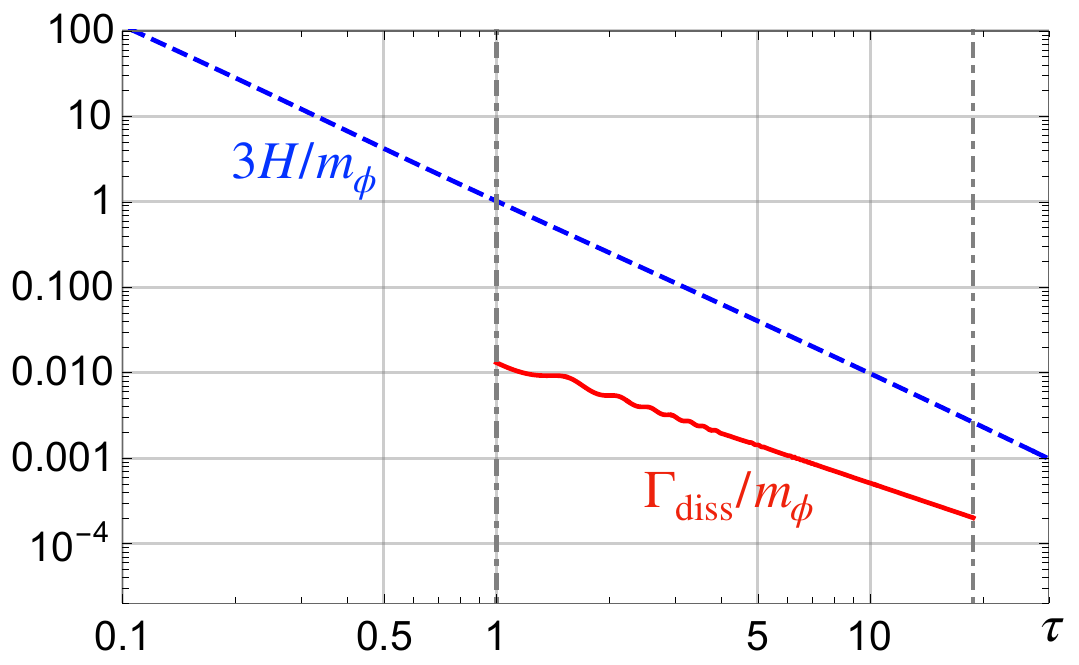}
\centering
\includegraphics[width=7.5cm]{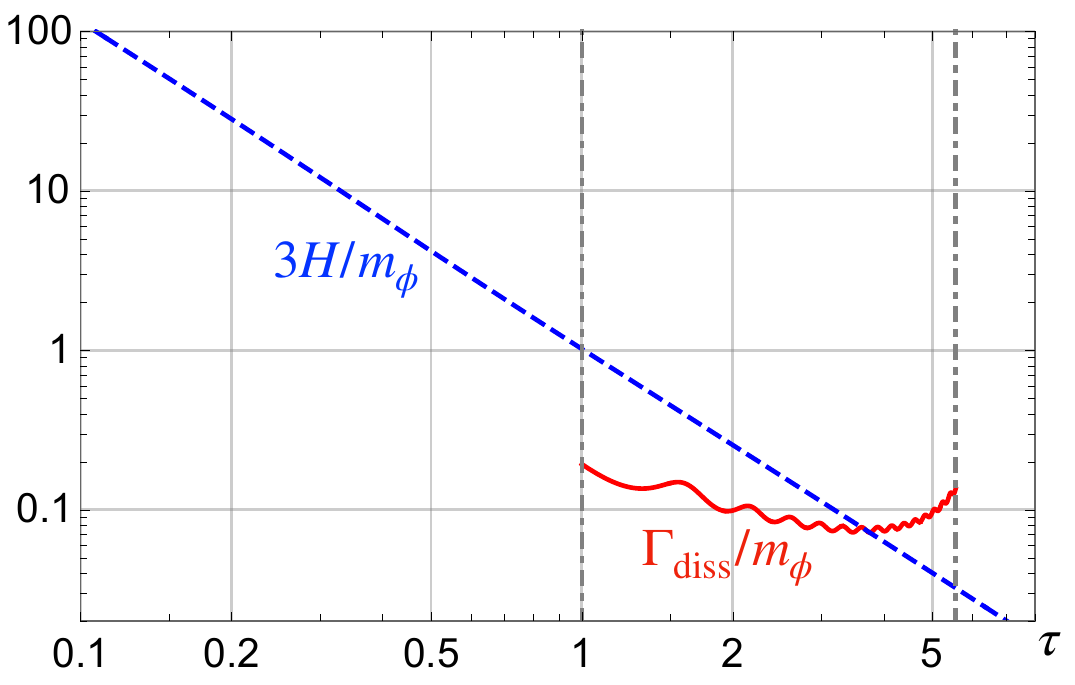}
\centering
\caption{Dissipation rate (red solid) and Hubble parameter (blue dashed) as a function of $\tau$, normalized by the axion mass $m_\phi$. The left vertical dot-dashed line represents the onset of the axion oscillation $(\tau_{\rm osc}=1)$, and the right one is the endpoint of the dissipation effect. 
}
\label{fig:hierarchy}
\end{figure}

Even if the dissipation rate is smaller than $3H$ at the onset of the axion oscillation, the dissipation term is not negligible at a later time, suppressing the axion abundance. 
To understand this, we compare $\Gamma_{\rm diss}/m_\phi$ (red solid curve) and $3H/m_\phi$ (blue dashed line) as a function of $\tau$ in \FIG{fig:hierarchy}.
The left vertical dot-dashed line represents the time at the onset of the axion oscillation $(\tau_{\rm ini}=1)$. 
The right vertical dot-dashed line is the endpoint of the dissipation effect that is determined by $T = T_{\rm end}$ (see \eq{endpoint}).
The dissipation rate decreases with time more slowly than the Hubble parameter, as we expected. 
This is why the dissipation can be efficient at a later time, even if it is not strong at the onset of the oscillation.

%%%%%%%%%%%%%%%%%%%%%%%%%%%%%%%%%%%%%%%%%%%%%%%%%%%%%%%%%%%
%%%%%%%%%%%%%%%%%%%%%%%%%%%%%%%%%%%%%%%%%%%%%%%%%%%%%%%%%%%
\subsection{Axion abundance
\label{sec:abundance}}
Let us calculate the abundance of axion both analytically and numerically.
Using Eq.~(\ref{approx}), we obtain an analytical formula for the resulting axion energy density.
After the axion starts to oscillate with an initial amplitude $\theta_{\rm ini}f_\phi$, 
the solution is given by
\beq
\rho_\phi =\lmk\frac{\tau}{\tau_{\rm osc}}\rmk^{-3}\rho_{\phi, {\rm ini}}\left[1-3\sqrt{2}\frac{\Gamma_{\rm diss}(\tau_{\rm osc})}{m_\phi}(\sqrt{\tau}-\sqrt{\tau_{\rm osc}})\right]^2,
\label{rho}
\eeq
where $\rho_{\phi,{\rm ini}}$ $(\simeq m_\phi^2f_\phi^2\theta_{\rm ini}^2/2)$ is the axion energy density at the onset of the oscillation, and we have neglected the time dependence of $g_*$ and $g_{*s}$ for simplicity. 
$\tau_{\rm osc}$ ($\equiv 1$) represents the timing at the onset of the axion oscillation, which is determined by $3H(\tau_{\rm osc})=m_\phi$.
When $\tau=\tau_{\rm end}$ at which $T= T_{\rm end}$ or $\rho_\phi = m_\phi^2\delta\phi_{\rm first}^2/2$, the bracket leads to the suppression factor of 
\beq
\Delta_{\rm SF} &\equiv& \left.\frac{\rho_\phi(\Omega_M\neq0)}{\rho_\phi(\Omega_M=0)}\right|_{\tau=\tau_{\rm end}}
\label{SFdef}
\\
&=& \left[1-3\sqrt{2}\frac{\Gamma_{\rm diss}(\tau_{\rm osc})}{m_\phi}(\sqrt{\tau_{\rm end}}-\sqrt{\tau_{\rm osc}})\right]^2,
\label{SFdef2}
\eeq
where we have used \eq{rho}.
The suppression factor can be alternatively written as 
\beq
\label{sup}
\Delta_{\rm SF} 
\simeq \lmk\frac{\tau_{\rm end}}{\tau_{\rm osc}}\rmk^3\lmk\frac{\delta\phi_{\rm first}}{\phi_{\rm ini}}\rmk^2,
\eeq
where we have used \eq{endpoint} and $\left. \rho_\phi(\Omega_M=0)\right|_{\tau=\tau_{\rm end}} \simeq m_\phi^2 \phi_{\rm ini}^2/2 \times (\tau_{\rm osc} / \tau_{\rm end})^3$. 
Therefore 
the suppression factor is determined by the ratio $\delta\phi_{\rm first}/\phi_{\rm ini}$ and the endpoint $\tau_{\rm end}$. 
The latter can be estimated numerically.
Note that the suppression factor is equal to unity when the inequality (\ref{Tendlow}) is saturated.

\begin{figure}[t!]
\includegraphics[width=7.6cm]{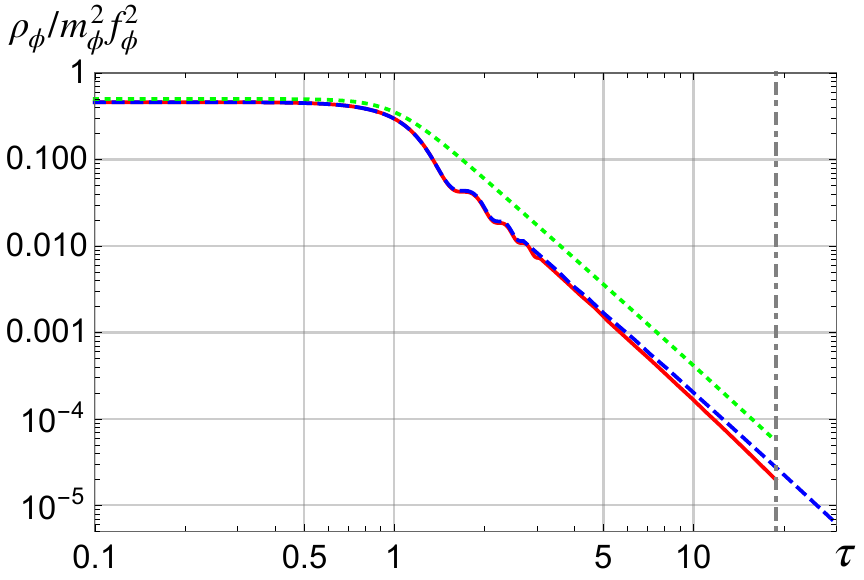}
\centering
\includegraphics[width=7.6cm]{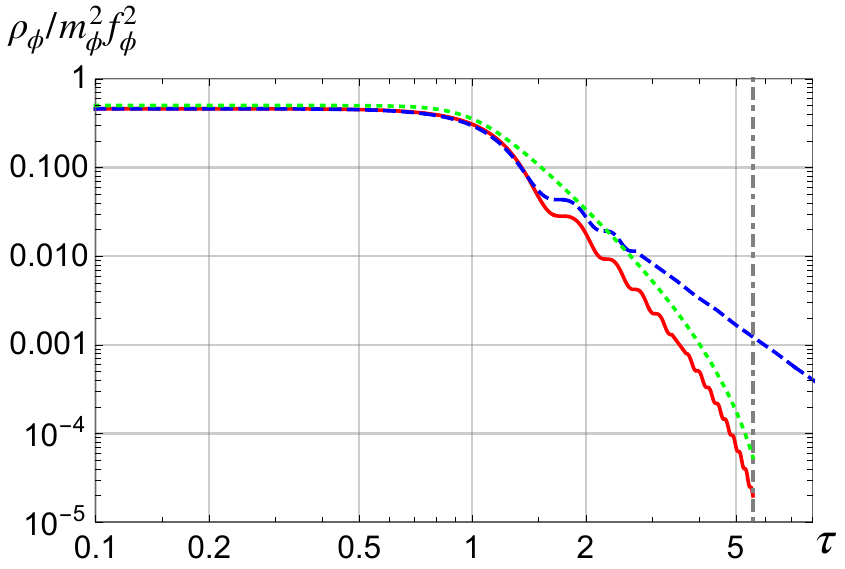}
\centering
\caption{The time evolution of the axion energy density as a function of $\tau$. The parameters are taken the same with \FIG{fig:dynamics}.
The green dotted line denotes the analytical result (\ref{rho}).
The gray dot-dashed line is the endpoint of the dissipation effect.}
\label{fig:density}
\end{figure}

We illustrate the time evolution of the energy density in \FIG{fig:density}. The parameters are taken as the same values with those used in \FIG{fig:dynamics}.
The red solid and blue dashed line respectively denote the numerical results with and without the dissipation effect, which are estimated from Eq.~(\ref{eom}).
The green dotted line denotes the analytical result (\ref{rho}), where we have set $\tau_{\rm osc}=1$.
The evolution of the axion energy density does not seem to change in the left panel. 
On the other hand, the energy density is significantly suppressed due to the dissipation effect in the right panel. We can see that the analytic estimate is consistent with the numerical one.

\begin{figure}[t!]
\includegraphics[width=11cm]{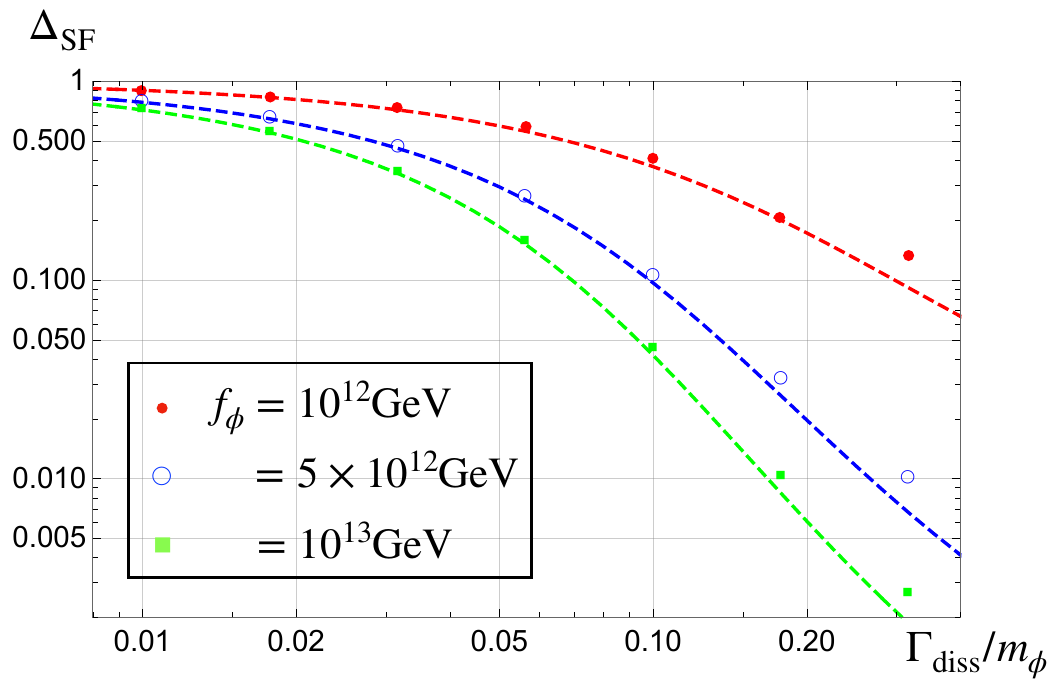}
\centering
\caption{Suppression factor as a function of the dissipation rate
at the onset of oscillations. 
We take $f_H=10^8{\rm GeV}$, $m_\phi=1\mu{\rm eV}$, $\alpha_H=10^{-2}$, and $\theta_{\rm ini}=1$. The red bullet, blue circle, and green square denote the results for $f_\phi=10^{12}, 5\times10^{12}, 10^{13}\GeV$, respectively.
The dashed lines represent the semi-analytic estimate based on Eq.~(\ref{SFdef2}) or equivalently (\ref{sup}), 
where $\tau_{\rm end}$ is computed numerically in terms of (\ref{rho}).
}
\label{fig:SF}
\end{figure}

The suppression factor, defined by \eq{SFdef}, is also numerically calculated as a function of the dissipation rate $\Gamma_{\rm diss}(\tau_{\rm osc})/m_\phi$ in \FIG{fig:SF}. 
We take $f_H=10^8{\rm GeV}$, $m_\phi=1\mu{\rm eV}$, $\alpha_H=10^{-2}$, and $\theta_{\rm ini}=1$. The red bullet, blue circle, and green square denote the results for $f_\phi=10^{12}$, $5\times10^{12}$, and $10^{13}\GeV$, respectively.
One can see that the abundance is suppressed more strongly for a lower value of $\delta\phi_{\rm first}/\phi_{\rm ini}$ ($\simeq 2 \pi \alpha_H^{-1} f_H / f_\phi$). 
In addition, we show the semi-analytic results which are represented by each dashed line in \FIG{fig:SF}. 
We can numerically estimate the endpoint $T_{\rm end}$ of the dissipation effect by equating \eq{rho} at $T=T_{\rm end}$, and the suppression factor is depicted by using the results and Eq.~(\ref{SFdef2}) or equivalently \eq{sup}.

%%%%%%%%%%%%%%%%%%%%%%%%%%%%%%%%%%%%%%%%%%%%%%%%%%%%%%%%%%%
%%%%%%%%%%%%%%%%%%%%%%%%%%%%%%%%%%%%%%%%%%%%%%%%%%%%%%%%%%%
\section{Discussion and conclusions
\label{sec:conclusion}}
In this paper, we have studied a new dissipative effect that has not been studied before with respect to the Witten effect of hidden sector monopoles on the axion.
This dissipative effect is caused by the $W$-boson pair production near the surface of dyons due to the Schwinger effect when the monopole charge becomes very large during the axion evolution.
As a result, the potential due to the Witten effect changes from time to time, and the amount of axion abundance that would be adiabatically suppressed in the absence of the Schwinger effect is enhanced. Alternatively, if the comparison is made with the usual misalignment mechanism in which the Witten and Schwinger effects are absent, the axion abundance is reduced due to dissipative effects.

We have discussed the details of this dissipation process under some approximations that help to numerically evaluate its effect on the axion dynamics. 
One of them is to ignore the effect of the potential generated by the Witten effect, which gradually changes with time due to the Schwinger effect, on the axion dynamics. We have confirmed that this is a reasonable approximation, at least for the parameter values we have adopted, but there is room for improvement in the description of the axion dynamics in other general parameter ranges, at least in terms of numerically computable approximations. Also, we have not studied in detail how the structure of the monopole solution is modified when the electric charge becomes large enough for the monopoles (or dyons) to decay and emit $W$-bosons via the Schwinger effect. While we have estimated how the monopole radius expands based on the dimensional argument, it would be important to study the internal structure, which allows us to determine more precisely when the dissipation becomes relevant.

In the present analysis we have also neglected a tachyonic preheating of dark photons, which becomes efficient when the axion has a moderately strong coupling to dark photons~\cite{Garretson:1992vt}. By the tachyonic production of dark photons, the axion abundance can be reduced to a few percent~\cite{Kitajima:2017peg} (see also Ref.~\cite{Agrawal:2017eqm}).  Note also that it takes some time after the onset of axion oscillations for the tachyonic preheating to be efficient, because the dark photons in the instability band must be sufficiently accumulated.
During and after the tachyonic production, the dissipation process is expected to occur via the Schwinger effect, but it is necessary to consider both effects to give a quantitative estimate of how much the axion abundance is reduced.

Finally, for simplicity, instead of the QCD axion, we have considered the case of the axion whose potential is constant in time. It is worthwhile to apply the new dissipation process to the case of the QCD axion.

%%%%%%%%%%%%%%%%%%%%%%%%%%%%%%%%%%%%%%%%%%%%%%%%%%%%%%%%%%%
%%%%%%%%%%%%%%%%%%%%%%%%%%%%%%%%%%%%%%%%%%%%%%%%%%%%%%%%%%%
\section*{Acknowledgments}
SN thanks Pusan National University and Institute for Basic Science for their hospitality during the visit, where part of this work was done.
The present work is supported by 
the Graduate Program on Physics for the Universe of Tohoku University (S.N.), JST SPRING, Grant Number JPMJSP2114 (S.N.),  
JSPS KAKENHI Grant Numbers 20H01894 (F.T.), 20H05851 (F.T. and M.Y.), and 23K13092 (M.Y.),
and also by the National Research Foundation (NRF) of Korea grant funded by the Korea government: Grants Numbers 2021R1A4A5031460 (K.S.J.) and RS-2023-00249330 (K.S.J.).
MY is supported by MEXT Leading Initiative for Excellent Young Researchers.  This article is based upon work from COST Action COSMIC WISPers CA21106,  supported by COST (European Cooperation in Science and Technology).

\appendix
\section{Electric dipole radiation via the Witten effect}
\label{sec:appendix}
In this Appendix, we discuss another origin of the dissipation effect induced by electric dipole radiation from a monopole-anti-monopole pair due to the Witten effect.

Here we consider monopole-anti-monopole pairs, ignoring the influence of the $W$ bosons. This corresponds to the case where the formation of the $W$ boson is suppressed or absent in the theory, or where the distance between the monopole and the anti-monopole is sufficiently small. Suppose that the monopoles in the background generically has a constant electric charge and an oscillating part due to the axion oscillations. We focus on the oscillating part since the constant dipole does not radiate dark photons. 
The electric charges oscillate due to the oscillating axion background, $\phi(t) = \phi_{\rm osc} \cos(m_\phi t)$.

The typical distance between the monopole and the nearest anti-monopole is bounded from above as $n_M^{-1/3}$, where $n_M$ is the monopole number density.
To be conservative,  let us assume for the moment that the bound is saturated, i.e., the monopoles and anti-monopoles are randomly distributed in space.  
The electric dipole moment of a monopole-anti-monopole pair separeted by the distance of $n_M^{-1/3}$ 
is given by
\begin{equation}
 \vec{p}_0 = n_M^{-\frac{1}{3}} e_H \frac{\phi_{\rm osc}}{2 \pi f_H}\frac{\vec{p}}{|\vec{p}|}.
\end{equation}
The total power radiated by the oscillating electric dipole moment $\vec{p}=\vec{p}_0\cos(\omega t)$ is~\cite{Jackson:1998nia}
\begin{equation}
     P(t) = \frac{\omega^4}{6\pi}  \vec{p}_0^2 \cos^2(kr-\omega t),
\end{equation}
where $\omega$ is the angular frequency, and  we use the Lorentz-Heaviside natural units.
If we take the time average for a fixed radius $r$, the $\cos^2$ function can be replaced with $1/2$.
Considering that there are roughly $n_M/2$ pairs of monopole and anti-monopole in the unit volume, the dark photon production rate is given by
\begin{equation}
    \frac{d\rho_\gamma}{dt}= \frac{\alpha_H m_\phi^4}{12 \pi^2} n_M^\frac{1}{3} \frac{\phi(t)^2}{f_H^2}.
    \label{eq:rhogammadis}
\end{equation}
Taking the time average over the oscillation period, we can
replace $m_\phi^2 \phi^2(t)$ with $\dot{\phi}^2(t)$ and rewrite the dissipation of \eq{eq:rhogammadis} as a frictional force on the axion. Then, the dissipation rate $\Gamma_{\rm ED}$ from this effect is
\begin{equation}
    \Gamma_{\rm ED} = \frac{\alpha_H m_\phi^2 n_M^\frac{1}{3}}{12 \pi^2 f_H^2}.
\end{equation}

The dissipation rate decrease as $a^{-1}$, while the Hubble parameter scales as $a^{-2}$ in the radiation dominated era, and so, in principle the dissipation rate can be important at a later time. 
By equating $\Gamma_{\rm ED}$ to the Hubble parameter in the radiation dominated Universe, 
we obtain the dissipation temperature of 
\begin{equation}
    T_{\rm ED} \simeq 0.34 {\rm GeV} \alpha_H  \left(\frac{g_*(T_{\rm ED})}{10^2}\right)^{-\frac{1}{6}}
    \left(\frac{m_\phi}{{\rm GeV}}\right)^2
     \left(\frac{\Omega_Mh^2}{10^{-3}}\right)^\frac{1}{3}
      \left(\frac{\rm TeV}{M_M}\right)^\frac{1}{3}
       \left(\frac{10^6{\rm GeV}}{f_H}\right)^2.
\end{equation}
At this temperature, 
the axion energy completely dissipates into dark photons via the dipole emission.

Note however that the formula for the electric dipole radiation is valid in the regime of the long wavelength compared to the distance between the monopole and anti-monopole. So we require 
\begin{equation}
\label{req}
    n_M (T_{\rm ED}) > m_\phi^3,
\end{equation}
or equivalently,
\begin{equation}
\label{req2}
    T_{\rm ED} > 2 \times 10^4 {\rm GeV}  \left(\frac{m_\phi}{{\rm GeV}}\right) \left(\frac{g_{*s}(T_{\rm ED})}{10^2}\right)^{-\frac{1}{3}}
     \left(\frac{\Omega_Mh^2}{10^{-3}}\right)^{-\frac{1}{3}}
     \left(\frac{M_M}{\rm TeV}\right)^\frac{1}{3}.
\end{equation}
Note that the perturbative decay rate of the axion into dark photons is comparable to $\Gamma_{\rm ED}$ when the condition of (\ref{req}) is saturated.

In order to satisfy the requirement (\ref{req}) or (\ref{req2}), we need to, e.g., increase the axion mass
by more than about $5$ orders of magnitude. 
Therefore, this dissipation process is unlikely to be important for light axions in a cosmological setting, unless the monopole density is temporally very large in the early universe, or there are many monopole-anti-monopole pairs whose distance is much shorter than $n_M^{-1/3}$. 
For example, monopole-anti-monopole bound states (called monopoliums) in Kaluza-Klein theories are known to be stable and their typical distances are very short~\cite{Vento:2020vsq}, and so, 
it would be interesting to study the electric dipole emission from them in
the presence of an oscillating axion field. In particular, the requirement (\ref{req}) is significantly relaxed due to the short distance between the monopole and the anti-monopole. Aslo, the dissipation rate becomes constant and can be much larger than the perturbative decay rate of the axion. Then, the axion condensate could evaporate faster than usual due to the electric dipole emission from the monopoliums.
We leave the detailed analysis for a future work.

\bibliography{reference}

\providecommand{\href}[2]{#2}\begingroup\raggedright\begin{thebibliography}{10}

\bibitem{Peccei:1977hh}
R.~D. Peccei and H.~R. Quinn, \emph{{CP Conservation in the Presence of
  Instantons}}, \href{https://doi.org/10.1103/PhysRevLett.38.1440}{\emph{Phys.
  Rev. Lett.} {\bfseries 38} (1977) 1440}.

\bibitem{Peccei:1977ur}
R.~D. Peccei and H.~R. Quinn, \emph{{Constraints Imposed by CP Conservation in
  the Presence of Instantons}},
  \href{https://doi.org/10.1103/PhysRevD.16.1791}{\emph{Phys. Rev. D}
  {\bfseries 16} (1977) 1791}.

\bibitem{Weinberg:1977ma}
S.~Weinberg, \emph{{A New Light Boson?}},
  \href{https://doi.org/10.1103/PhysRevLett.40.223}{\emph{Phys. Rev. Lett.}
  {\bfseries 40} (1978) 223}.

\bibitem{Wilczek:1977pj}
F.~Wilczek, \emph{{Problem of Strong $P$ and $T$ Invariance in the Presence of
  Instantons}}, \href{https://doi.org/10.1103/PhysRevLett.40.279}{\emph{Phys.
  Rev. Lett.} {\bfseries 40} (1978) 279}.

\bibitem{Jaeckel:2010ni}
J.~Jaeckel and A.~Ringwald, \emph{{The Low-Energy Frontier of Particle
  Physics}},
  \href{https://doi.org/10.1146/annurev.nucl.012809.104433}{\emph{Ann. Rev.
  Nucl. Part. Sci.} {\bfseries 60} (2010) 405}
  [\href{https://arxiv.org/abs/1002.0329}{{\ttfamily 1002.0329}}].

\bibitem{Ringwald:2012hr}
A.~Ringwald, \emph{{Exploring the Role of Axions and Other WISPs in the Dark
  Universe}}, \href{https://doi.org/10.1016/j.dark.2012.10.008}{\emph{Phys.
  Dark Univ.} {\bfseries 1} (2012) 116}
  [\href{https://arxiv.org/abs/1210.5081}{{\ttfamily 1210.5081}}].

\bibitem{Arias:2012az}
P.~Arias, D.~Cadamuro, M.~Goodsell, J.~Jaeckel, J.~Redondo and A.~Ringwald,
  \emph{{WISPy Cold Dark Matter}},
  \href{https://doi.org/10.1088/1475-7516/2012/06/013}{\emph{JCAP} {\bfseries
  06} (2012) 013} [\href{https://arxiv.org/abs/1201.5902}{{\ttfamily
  1201.5902}}].

\bibitem{Graham:2015ouw}
P.~W. Graham, I.~G. Irastorza, S.~K. Lamoreaux, A.~Lindner and K.~A. van
  Bibber, \emph{{Experimental Searches for the Axion and Axion-Like
  Particles}},
  \href{https://doi.org/10.1146/annurev-nucl-102014-022120}{\emph{Ann. Rev.
  Nucl. Part. Sci.} {\bfseries 65} (2015) 485}
  [\href{https://arxiv.org/abs/1602.00039}{{\ttfamily 1602.00039}}].

\bibitem{Marsh:2015xka}
D.~J.~E. Marsh, \emph{{Axion Cosmology}},
  \href{https://doi.org/10.1016/j.physrep.2016.06.005}{\emph{Phys. Rept.}
  {\bfseries 643} (2016) 1} [\href{https://arxiv.org/abs/1510.07633}{{\ttfamily
  1510.07633}}].

\bibitem{Irastorza:2018dyq}
I.~G. Irastorza and J.~Redondo, \emph{{New experimental approaches in the
  search for axion-like particles}},
  \href{https://doi.org/10.1016/j.ppnp.2018.05.003}{\emph{Prog. Part. Nucl.
  Phys.} {\bfseries 102} (2018) 89}
  [\href{https://arxiv.org/abs/1801.08127}{{\ttfamily 1801.08127}}].

\bibitem{DiLuzio:2020wdo}
L.~Di~Luzio, M.~Giannotti, E.~Nardi and L.~Visinelli, \emph{{The landscape of
  QCD axion models}},
  \href{https://doi.org/10.1016/j.physrep.2020.06.002}{\emph{Phys. Rept.}
  {\bfseries 870} (2020) 1} [\href{https://arxiv.org/abs/2003.01100}{{\ttfamily
  2003.01100}}].

\bibitem{Preskill:1982cy}
J.~Preskill, M.~B. Wise and F.~Wilczek, \emph{{Cosmology of the Invisible
  Axion}}, \href{https://doi.org/10.1016/0370-2693(83)90637-8}{\emph{Phys.
  Lett. B} {\bfseries 120} (1983) 127}.

\bibitem{Abbott:1982af}
L.~F. Abbott and P.~Sikivie, \emph{{A Cosmological Bound on the Invisible
  Axion}}, \href{https://doi.org/10.1016/0370-2693(83)90638-X}{\emph{Phys.
  Lett. B} {\bfseries 120} (1983) 133}.

\bibitem{Dine:1982ah}
M.~Dine and W.~Fischler, \emph{{The Not So Harmless Axion}},
  \href{https://doi.org/10.1016/0370-2693(83)90639-1}{\emph{Phys. Lett. B}
  {\bfseries 120} (1983) 137}.

\bibitem{Graham:2018jyp}
P.~W. Graham and A.~Scherlis, \emph{{Stochastic axion scenario}},
  \href{https://doi.org/10.1103/PhysRevD.98.035017}{\emph{Phys. Rev. D}
  {\bfseries 98} (2018) 035017}
  [\href{https://arxiv.org/abs/1805.07362}{{\ttfamily 1805.07362}}].

\bibitem{Guth:2018hsa}
F.~Takahashi, W.~Yin and A.~H. Guth, \emph{{QCD axion window and low-scale
  inflation}}, \href{https://doi.org/10.1103/PhysRevD.98.015042}{\emph{Phys.
  Rev. D} {\bfseries 98} (2018) 015042}
  [\href{https://arxiv.org/abs/1805.08763}{{\ttfamily 1805.08763}}].

\bibitem{Ho:2019ayl}
S.-Y. Ho, F.~Takahashi and W.~Yin, \emph{{Relaxing the Cosmological Moduli
  Problem by Low-scale Inflation}},
  \href{https://doi.org/10.1007/JHEP04(2019)149}{\emph{JHEP} {\bfseries 04}
  (2019) 149} [\href{https://arxiv.org/abs/1901.01240}{{\ttfamily
  1901.01240}}].

\bibitem{Nakagawa:2020eeg}
S.~Nakagawa, F.~Takahashi and W.~Yin, \emph{{Stochastic Axion Dark Matter in
  Axion Landscape}},
  \href{https://doi.org/10.1088/1475-7516/2020/05/004}{\emph{JCAP} {\bfseries
  05} (2020) 004} [\href{https://arxiv.org/abs/2002.12195}{{\ttfamily
  2002.12195}}].

\bibitem{Reig:2021ipa}
M.~Reig, \emph{{The stochastic axiverse}},
  \href{https://doi.org/10.1007/JHEP09(2021)207}{\emph{JHEP} {\bfseries 09}
  (2021) 207} [\href{https://arxiv.org/abs/2104.09923}{{\ttfamily
  2104.09923}}].

\bibitem{Murai:2023xjn}
K.~Murai, F.~Takahashi and W.~Yin, \emph{{The QCD Axion: A Unique Player in the
  Axiverse with Mixings}},  \href{https://arxiv.org/abs/2305.18677}{{\ttfamily
  2305.18677}}.

\bibitem{Co:2018mho}
R.~T. Co, E.~Gonzalez and K.~Harigaya, \emph{{Axion Misalignment Driven to the
  Hilltop}}, \href{https://doi.org/10.1007/JHEP05(2019)163}{\emph{JHEP}
  {\bfseries 05} (2019) 163}
  [\href{https://arxiv.org/abs/1812.11192}{{\ttfamily 1812.11192}}].

\bibitem{Takahashi:2019pqf}
F.~Takahashi and W.~Yin, \emph{{QCD axion on hilltop by a phase shift of
  $\pi$}}, \href{https://doi.org/10.1007/JHEP10(2019)120}{\emph{JHEP}
  {\bfseries 10} (2019) 120}
  [\href{https://arxiv.org/abs/1908.06071}{{\ttfamily 1908.06071}}].

\bibitem{Huang:2020etx}
J.~Huang, A.~Madden, D.~Racco and M.~Reig, \emph{{Maximal axion misalignment
  from a minimal model}},
  \href{https://doi.org/10.1007/JHEP10(2020)143}{\emph{JHEP} {\bfseries 10}
  (2020) 143} [\href{https://arxiv.org/abs/2006.07379}{{\ttfamily
  2006.07379}}].

\bibitem{Agrawal:2017eqm}
P.~Agrawal, G.~Marques-Tavares and W.~Xue, \emph{{Opening up the QCD axion
  window}}, \href{https://doi.org/10.1007/JHEP03(2018)049}{\emph{JHEP}
  {\bfseries 03} (2018) 049}
  [\href{https://arxiv.org/abs/1708.05008}{{\ttfamily 1708.05008}}].

\bibitem{Kitajima:2017peg}
N.~Kitajima, T.~Sekiguchi and F.~Takahashi, \emph{{Cosmological abundance of
  the QCD axion coupled to hidden photons}},
  \href{https://doi.org/10.1016/j.physletb.2018.04.024}{\emph{Phys. Lett. B}
  {\bfseries 781} (2018) 684}
  [\href{https://arxiv.org/abs/1711.06590}{{\ttfamily 1711.06590}}].

\bibitem{Kitajima:2023pby}
N.~Kitajima and F.~Takahashi, \emph{{Resonant production of dark photons from
  axions without a large coupling}},
  \href{https://doi.org/10.1103/PhysRevD.107.123518}{\emph{Phys. Rev. D}
  {\bfseries 107} (2023) 123518}
  [\href{https://arxiv.org/abs/2303.05492}{{\ttfamily 2303.05492}}].

\bibitem{Higaki:2016yqk}
T.~Higaki, K.~S. Jeong, N.~Kitajima and F.~Takahashi, \emph{{Quality of the
  Peccei-Quinn symmetry in the Aligned QCD Axion and Cosmological
  Implications}}, \href{https://doi.org/10.1007/JHEP06(2016)150}{\emph{JHEP}
  {\bfseries 06} (2016) 150}
  [\href{https://arxiv.org/abs/1603.02090}{{\ttfamily 1603.02090}}].

\bibitem{Jeong:2022kdr}
K.~S. Jeong, K.~Matsukawa, S.~Nakagawa and F.~Takahashi, \emph{{Cosmological
  effects of Peccei-Quinn symmetry breaking on QCD axion dark matter}},
  \href{https://doi.org/10.1088/1475-7516/2022/03/026}{\emph{JCAP} {\bfseries
  03} (2022) 026} [\href{https://arxiv.org/abs/2201.00681}{{\ttfamily
  2201.00681}}].

\bibitem{Nakagawa:2022wwm}
S.~Nakagawa, F.~Takahashi, M.~Yamada and W.~Yin, \emph{{Axion dark matter from
  first-order phase transition, and very high energy photons from GRB
  221009A}}, \href{https://doi.org/10.1016/j.physletb.2023.137824}{\emph{Phys.
  Lett. B} {\bfseries 839} (2023) 137824}
  [\href{https://arxiv.org/abs/2210.10022}{{\ttfamily 2210.10022}}].

\bibitem{Papageorgiou:2022prc}
A.~Papageorgiou, P.~Qu\'\i{}lez and K.~Schmitz, \emph{{Axion dark matter from
  frictional misalignment}},
  \href{https://doi.org/10.1007/JHEP01(2023)169}{\emph{JHEP} {\bfseries 01}
  (2023) 169} [\href{https://arxiv.org/abs/2206.01129}{{\ttfamily
  2206.01129}}].

\bibitem{Choi:2022nlt}
K.~Choi, S.~H. Im, H.~J. Kim and H.~Seong, \emph{{Axion dark matter with
  thermal friction}},
  \href{https://doi.org/10.1007/JHEP02(2023)180}{\emph{JHEP} {\bfseries 02}
  (2023) 180} [\href{https://arxiv.org/abs/2206.01462}{{\ttfamily
  2206.01462}}].

\bibitem{Witten:1979ey}
E.~Witten, \emph{{Dyons of Charge e theta/2 pi}},
  \href{https://doi.org/10.1016/0370-2693(79)90838-4}{\emph{Phys. Lett. B}
  {\bfseries 86} (1979) 283}.

\bibitem{Fischler:1983sc}
W.~Fischler and J.~Preskill, \emph{{DYON - AXION DYNAMICS}},
  \href{https://doi.org/10.1016/0370-2693(83)91260-1}{\emph{Phys. Lett. B}
  {\bfseries 125} (1983) 165}.

\bibitem{Kawasaki:2015lpf}
M.~Kawasaki, F.~Takahashi and M.~Yamada, \emph{{Suppressing the QCD Axion
  Abundance by Hidden Monopoles}},
  \href{https://doi.org/10.1016/j.physletb.2015.12.075}{\emph{Phys. Lett. B}
  {\bfseries 753} (2016) 677}
  [\href{https://arxiv.org/abs/1511.05030}{{\ttfamily 1511.05030}}].

\bibitem{Nomura:2015xil}
Y.~Nomura, S.~Rajendran and F.~Sanches, \emph{{Axion Isocurvature and Magnetic
  Monopoles}},
  \href{https://doi.org/10.1103/PhysRevLett.116.141803}{\emph{Phys. Rev. Lett.}
  {\bfseries 116} (2016) 141803}
  [\href{https://arxiv.org/abs/1511.06347}{{\ttfamily 1511.06347}}].

\bibitem{Kawasaki:2017xwt}
M.~Kawasaki, F.~Takahashi and M.~Yamada, \emph{{Adiabatic suppression of the
  axion abundance and isocurvature due to coupling to hidden monopoles}},
  \href{https://doi.org/10.1007/JHEP01(2018)053}{\emph{JHEP} {\bfseries 01}
  (2018) 053} [\href{https://arxiv.org/abs/1708.06047}{{\ttfamily
  1708.06047}}].

\bibitem{Nakagawa:2020zjr}
S.~Nakagawa, F.~Takahashi and M.~Yamada, \emph{{Trapping Effect for QCD Axion
  Dark Matter}},
  \href{https://doi.org/10.1088/1475-7516/2021/05/062}{\emph{JCAP} {\bfseries
  05} (2021) 062} [\href{https://arxiv.org/abs/2012.13592}{{\ttfamily
  2012.13592}}].

\bibitem{Heisenberg:1936nmg}
W.~Heisenberg and H.~Euler, \emph{{Consequences of Dirac's theory of
  positrons}}, \href{https://doi.org/10.1007/BF01343663}{\emph{Z. Phys.}
  {\bfseries 98} (1936) 714}
  [\href{https://arxiv.org/abs/physics/0605038}{{\ttfamily physics/0605038}}].

\bibitem{Schwinger:1951nm}
J.~S. Schwinger, \emph{{On gauge invariance and vacuum polarization}},
  \href{https://doi.org/10.1103/PhysRev.82.664}{\emph{Phys. Rev.} {\bfseries
  82} (1951) 664}.

\bibitem{tHooft:1974kcl}
G.~'t~Hooft, \emph{{Magnetic Monopoles in Unified Gauge Theories}},
  \href{https://doi.org/10.1016/0550-3213(74)90486-6}{\emph{Nucl. Phys. B}
  {\bfseries 79} (1974) 276}.

\bibitem{Polyakov:1974ek}
A.~M. Polyakov, \emph{{Particle Spectrum in Quantum Field Theory}}, {\emph{JETP
  Lett.} {\bfseries 20} (1974) 194}.

\bibitem{Kibble:1976sj}
T.~W.~B. Kibble, \emph{{Topology of Cosmic Domains and Strings}},
  \href{https://doi.org/10.1088/0305-4470/9/8/029}{\emph{J. Phys. A} {\bfseries
  9} (1976) 1387}.

\bibitem{Zurek:1985qw}
W.~H. Zurek, \emph{{Cosmological Experiments in Superfluid Helium?}},
  \href{https://doi.org/10.1038/317505a0}{\emph{Nature} {\bfseries 317} (1985)
  505}.

\bibitem{Baek:2013dwa}
S.~Baek, P.~Ko and W.-I. Park, \emph{{Hidden sector monopole, vector dark
  matter and dark radiation with Higgs portal}},
  \href{https://doi.org/10.1088/1475-7516/2014/10/067}{\emph{JCAP} {\bfseries
  10} (2014) 067} [\href{https://arxiv.org/abs/1311.1035}{{\ttfamily
  1311.1035}}].

\bibitem{Khoze:2014woa}
V.~V. Khoze and G.~Ro, \emph{{Dark matter monopoles, vectors and photons}},
  \href{https://doi.org/10.1007/JHEP10(2014)061}{\emph{JHEP} {\bfseries 10}
  (2014) 061} [\href{https://arxiv.org/abs/1406.2291}{{\ttfamily 1406.2291}}].

\bibitem{Fuentes-Martin:2019bue}
J.~Fuentes-Mart\'\i{}n, M.~Reig and A.~Vicente, \emph{{Strong $CP$ problem with
  low-energy emergent QCD: The 4321 case}},
  \href{https://doi.org/10.1103/PhysRevD.100.115028}{\emph{Phys. Rev. D}
  {\bfseries 100} (2019) 115028}
  [\href{https://arxiv.org/abs/1907.02550}{{\ttfamily 1907.02550}}].

\bibitem{Csaki:2019vte}
C.~Cs\'aki, M.~Ruhdorfer and Y.~Shirman, \emph{{UV Sensitivity of the Axion
  Mass from Instantons in Partially Broken Gauge Groups}},
  \href{https://doi.org/10.1007/JHEP04(2020)031}{\emph{JHEP} {\bfseries 04}
  (2020) 031} [\href{https://arxiv.org/abs/1912.02197}{{\ttfamily
  1912.02197}}].

\bibitem{Buen-Abad:2019uoc}
M.~A. Buen-Abad and J.~Fan, \emph{{Dynamical axion misalignment with small
  instantons}}, \href{https://doi.org/10.1007/JHEP12(2019)161}{\emph{JHEP}
  {\bfseries 12} (2019) 161}
  [\href{https://arxiv.org/abs/1911.05737}{{\ttfamily 1911.05737}}].

\bibitem{Kim:2004rp}
J.~E. Kim, H.~P. Nilles and M.~Peloso, \emph{{Completing natural inflation}},
  \href{https://doi.org/10.1088/1475-7516/2005/01/005}{\emph{JCAP} {\bfseries
  01} (2005) 005} [\href{https://arxiv.org/abs/hep-ph/0409138}{{\ttfamily
  hep-ph/0409138}}].

\bibitem{Choi:2014rja}
K.~Choi, H.~Kim and S.~Yun, \emph{{Natural inflation with multiple
  sub-Planckian axions}},
  \href{https://doi.org/10.1103/PhysRevD.90.023545}{\emph{Phys. Rev. D}
  {\bfseries 90} (2014) 023545}
  [\href{https://arxiv.org/abs/1404.6209}{{\ttfamily 1404.6209}}].

\bibitem{Higaki:2014qua}
T.~Higaki, N.~Kitajima and F.~Takahashi, \emph{{Hidden axion dark matter
  decaying through mixing with QCD axion and the 3.5 keV X-ray line}},
  \href{https://doi.org/10.1088/1475-7516/2014/12/004}{\emph{JCAP} {\bfseries
  12} (2014) 004} [\href{https://arxiv.org/abs/1408.3936}{{\ttfamily
  1408.3936}}].

\bibitem{Higaki:2015jag}
T.~Higaki, K.~S. Jeong, N.~Kitajima and F.~Takahashi, \emph{{The QCD Axion from
  Aligned Axions and Diphoton Excess}},
  \href{https://doi.org/10.1016/j.physletb.2016.01.055}{\emph{Phys. Lett. B}
  {\bfseries 755} (2016) 13}
  [\href{https://arxiv.org/abs/1512.05295}{{\ttfamily 1512.05295}}].

\bibitem{Choi:2015fiu}
K.~Choi and S.~H. Im, \emph{{Realizing the relaxion from multiple axions and
  its UV completion with high scale supersymmetry}},
  \href{https://doi.org/10.1007/JHEP01(2016)149}{\emph{JHEP} {\bfseries 01}
  (2016) 149} [\href{https://arxiv.org/abs/1511.00132}{{\ttfamily
  1511.00132}}].

\bibitem{Kaplan:2015fuy}
D.~E. Kaplan and R.~Rattazzi, \emph{{Large field excursions and approximate
  discrete symmetries from a clockwork axion}},
  \href{https://doi.org/10.1103/PhysRevD.93.085007}{\emph{Phys. Rev. D}
  {\bfseries 93} (2016) 085007}
  [\href{https://arxiv.org/abs/1511.01827}{{\ttfamily 1511.01827}}].

\bibitem{Giudice:2016yja}
G.~F. Giudice and M.~McCullough, \emph{{A Clockwork Theory}},
  \href{https://doi.org/10.1007/JHEP02(2017)036}{\emph{JHEP} {\bfseries 02}
  (2017) 036} [\href{https://arxiv.org/abs/1610.07962}{{\ttfamily
  1610.07962}}].

\bibitem{Farina:2016tgd}
M.~Farina, D.~Pappadopulo, F.~Rompineve and A.~Tesi, \emph{{The photo-philic
  QCD axion}}, \href{https://doi.org/10.1007/JHEP01(2017)095}{\emph{JHEP}
  {\bfseries 01} (2017) 095}
  [\href{https://arxiv.org/abs/1611.09855}{{\ttfamily 1611.09855}}].

\bibitem{Long:2018nsl}
A.~J. Long, \emph{{Cosmological Aspects of the Clockwork Axion}},
  \href{https://doi.org/10.1007/JHEP07(2018)066}{\emph{JHEP} {\bfseries 07}
  (2018) 066} [\href{https://arxiv.org/abs/1803.07086}{{\ttfamily
  1803.07086}}].

\bibitem{Chiang:2020aui}
C.-W. Chiang and B.-Q. Lu, \emph{{Testing clockwork axion with gravitational
  waves}}, \href{https://doi.org/10.1088/1475-7516/2021/05/049}{\emph{JCAP}
  {\bfseries 05} (2021) 049}
  [\href{https://arxiv.org/abs/2012.14071}{{\ttfamily 2012.14071}}].

\bibitem{Hawking:1974rv}
S.~W. Hawking, \emph{{Black hole explosions}},
  \href{https://doi.org/10.1038/248030a0}{\emph{Nature} {\bfseries 248} (1974)
  30}.

\bibitem{Hawking:1975vcx}
S.~W. Hawking, \emph{{Particle Creation by Black Holes}},
  \href{https://doi.org/10.1007/BF02345020}{\emph{Commun. Math. Phys.}
  {\bfseries 43} (1975) 199}.

\bibitem{Julia:1975ff}
B.~Julia and A.~Zee, \emph{{Poles with Both Magnetic and Electric Charges in
  Nonabelian Gauge Theory}},
  \href{https://doi.org/10.1103/PhysRevD.11.2227}{\emph{Phys. Rev. D}
  {\bfseries 11} (1975) 2227}.

\bibitem{Prasad:1975kr}
M.~K. Prasad and C.~M. Sommerfield, \emph{{An Exact Classical Solution for the
  't Hooft Monopole and the Julia-Zee Dyon}},
  \href{https://doi.org/10.1103/PhysRevLett.35.760}{\emph{Phys. Rev. Lett.}
  {\bfseries 35} (1975) 760}.

\bibitem{Smilga:1988pk}
A.~V. Smilga, \emph{{On scattering of W bosons on monopoles. (In Russian)}},
  {\emph{Sov. J. Nucl. Phys.} {\bfseries 47} (1988) 692}.

\bibitem{Linde:1996cx}
A.~D. Linde, \emph{{Relaxing the cosmological moduli problem}},
  \href{https://doi.org/10.1103/PhysRevD.53.R4129}{\emph{Phys. Rev. D}
  {\bfseries 53} (1996) R4129}
  [\href{https://arxiv.org/abs/hep-th/9601083}{{\ttfamily hep-th/9601083}}].

\bibitem{Saikawa:2018rcs}
K.~Saikawa and S.~Shirai, \emph{{Primordial gravitational waves, precisely: The
  role of thermodynamics in the Standard Model}},
  \href{https://doi.org/10.1088/1475-7516/2018/05/035}{\emph{JCAP} {\bfseries
  05} (2018) 035} [\href{https://arxiv.org/abs/1803.01038}{{\ttfamily
  1803.01038}}].

\bibitem{Garretson:1992vt}
W.~D. Garretson, G.~B. Field and S.~M. Carroll, \emph{{Primordial magnetic
  fields from pseudoGoldstone bosons}},
  \href{https://doi.org/10.1103/PhysRevD.46.5346}{\emph{Phys. Rev. D}
  {\bfseries 46} (1992) 5346}
  [\href{https://arxiv.org/abs/hep-ph/9209238}{{\ttfamily hep-ph/9209238}}].

\bibitem{Jackson:1998nia}
J.~D. Jackson, \emph{{Classical Electrodynamics}}. Wiley, 1998.

\bibitem{Vento:2020vsq}
V.~Vento, \emph{{Primordial monopolium as dark matter}},
  \href{https://doi.org/10.1140/epjc/s10052-021-09027-6}{\emph{Eur. Phys. J. C}
  {\bfseries 81} (2021) 229}
  [\href{https://arxiv.org/abs/2011.10327}{{\ttfamily 2011.10327}}].

\end{thebibliography}\endgroup

\end{document}